\documentclass[journal]{IEEEtran}
\ifCLASSINFOpdf
\else
\fi
\ifCLASSOPTIONcompsoc
 \usepackage[caption=false,font=normalsize,labelfont=sf,textfont=sf]{subfig}
\else
 \usepackage[caption=false,font=footnotesize]{subfig}
\fi

\usepackage{graphicx}
\usepackage{amsmath}
\usepackage{mathtools}
\usepackage{epsfig}
\usepackage{float}
\usepackage{lipsum}
\usepackage{stfloats}
\usepackage{array}
\usepackage{amssymb}
\usepackage{cite}
\usepackage{url}
\usepackage{algorithm2e}
\usepackage{algorithmic}
\usepackage{multirow}
\usepackage{fancyh}
\usepackage{flushend}
\usepackage{upgreek}
\usepackage{dsfont}
\usepackage{gensymb}
\usepackage{amsthm}
\usepackage{physics}

\usepackage{multicol, blindtext}
\usepackage{mathtools, cuted}

\usepackage{color}
\usepackage{bm}
\usepackage{booktabs}
\usepackage{tablefootnote}
\usepackage{threeparttable}
\usepackage{makecell}

\usepackage{tikz}
\usepackage{pgfmath, pgfplots}
\usetikzlibrary{3d, shapes.geometric, calc}
\usepgfplotslibrary{fillbetween}
\usepgfplotslibrary{groupplots}
\usepackage{color}

\usepackage[normalem]{ulem}

\pgfplotsset{table/search path={../results,results}}

\theoremstyle{remark}
\newtheorem{theorem}{Theorem}
\newtheorem{lemma}{Lemma}

\newtheorem{corollary}{Corollary}
\newtheorem{definition}{Definition}

\newtheorem{proposition}{Proposition}

\newtheorem{remark}{Remark}

\usepackage{pifont}
\DeclareMathOperator*{\argmin}{argmin} 

\def\one{{\boldsymbol{1}}}
\def\zero{{\boldsymbol{0}}}

\def\bb0{{\mathbb{0}}}

\def\ba{{\boldsymbol{a}}}
\def\bb{{\boldsymbol{b}}}

\def\bee{{\boldsymbol{e}}}

\def\bh{{\boldsymbol{h}}}

\def\bp{{\boldsymbol{p}}}
\def\bq{{\boldsymbol{q}}}
\def\br{{\boldsymbol{r}}}
\def\bs{{\boldsymbol{s}}}
\def\bt{{\boldsymbol{t}}}
\def\bu{{\boldsymbol{u}}}
\def\bv{{\boldsymbol{v}}}

\def\bx{{\boldsymbol{x}}}

\def\bz{{\boldsymbol{z}}}
\def\b0{{\boldsymbol{0}}}

\def\bE{{\boldsymbol{E}}}

\def\bG{{\boldsymbol{G}}}
\def\bH{{\boldsymbol{H}}}
\def\bI{{\boldsymbol{I}}}

\def\x{{\mathrm{x}}}

\def\b{{\mathrm{b}}}
\def\h{{\mathrm{h}}}

\def\k{{\mathrm{k}}}

\def\r0{{\mathbf{0}}}

\def\bbC{{\mathbb{C}}}

\def\bbE{{\mathbb{E}}}

\def\bbR{{\mathbb{R}}}

\def\bbZ{{\mathbb{Z}}}

\def\cG{\mathcal{G}}

\def\cT{\mathcal{T}}

\def\bPi{\bm \Pi}

\def\bsf0{{\bm{\mathsf{0}}}}

\def\N0{{N_{\mathrm{0}}}}

\def\j{\mathrm{j}}

\def\g{{g}}

\def\bsf{{\boldsymbol{s}_\mathrm{f}}}

\newcommand{\be}{\begin{equation}}
\newcommand{\ee}{\end{equation}}
\newcommand{\bal}{\begin{align}}
\newcommand{\eal}{\end{align}}

\def\SNR    {{\mathsf{SNR}}}

\normalsize
\begin{document}
%
\title{Channel Orthogonalization Via Movable Antennas
\thanks{H. Do and A. Lozano are with Univ. Pompeu Fabra, 08018 Barcelona (e-mail: \{heedong.do, angel.lozano\}@upf.edu).
Their work is supported by the Maria de Maeztu Units of Excellence Programme CEX2021-001195-M funded by MICIU/AEI/10.13039/501100011033, and by the Departament de Recerca i Universitats de la Generalitat de Catalunya.}
}

\author{\IEEEauthorblockN{Heedong~Do},
 {\it Member,~IEEE},
 \and
\IEEEauthorblockN{Angel~Lozano},
{\it Fellow,~IEEE}
}
\maketitle



%


\maketitle

\begin{abstract}
This paper investigates the orthogonalization of far-field multiuser channels through the optimal 
positioning of the antennas constituting an array.
To solve the ensuing continuous geometry optimization problem, a graph-theoretic framework is introduced that maps pairwise user channel correlations to graph edges, translating channel orthogonalization into a discrete graph-covering problem on a complete graph. The minimum number of array antennas required for orthogonalization is established for 1D, 2D, and 3D array geometries, and a binary linear program is formulated that identifies the most compact array attaining such orthogonality.
The graph framework is also shown to accommodate null-steering scenarios and multipath channels. Furthermore, a fundamental tradeoff is revealed whereby relaxing exact orthogonality to near-orthogonality reduces the antenna count at the expense of a less compact array, and a method is presented to enforce minimum antenna spacings without sacrificing orthogonality.
\end{abstract}



%
\IEEEpeerreviewmaketitle


\section{Introduction}

The idea of optimizing antenna positions for better utility---usually lower sidelobes and nulls in certain directions---has been extensively explored in the antenna community, mostly by means of evolutionary algorithms \cite{khodier2005linear, bevelacqua2007minimum, goudos2010application, lin2010synthesis, bhattacharya2012position, mahto2016novel};
in particular, the
optimization of antenna positions over a discrete grid is termed array thinning \cite{haupt1994thinned, keizer2009large, oliveri2009linear}. 
These methods are primarily developed for stationary environments where the array is optimized once and deployed without further real-time adjustments, yet in principle could adaptively optimize the array in response to environmental changes \cite{rocca2011interference, poli2013reconfigurable, haupt2015adaptively}.

The aforementioned treatments are meant for line-of-sight (LOS) channels. In the face of multipath propagation, the counterpart to array thinning is antenna selection---an old technique that evolved
from being driven by diversity against fading to being motivated by the improvement of the conditions for spatial multiplexing
\cite{917094,molisch2004mimo, sanayei2004antenna, dai2006optimal}.

Of late, there is a resurgence of interest in arranging array geometries to render the propagation more favorable for a given number of antennas or, equivalently, to reduce the number of antennas for the same level of channel conditioning \cite{bjornson2026antenna}.
This fresh impetus relates to new opportunities for reconfigurability such as those offered by dynamic metasurface antennas \cite{9324910},
mechanical or electronic actuators \cite{mirmozafari2021mechanically},
or swarms of cooperating drones forming an array \cite{mozaffari2019communications,11512076,wan2026movable}.
Depending on the form of implementation and the extent, the concept is referred to as movable antennas \cite{zhu2024movable}, reconfigurable antennas \cite{do2021reconfigurable,do2021rotatable}, 
rotatable antennas \cite{zheng2026rotatable}, fluid antennas \cite{wu2026fluid}, or pinching antennas \cite{yang2026pinching}. In this paper, for the sake of specificity, the term movable antennas is adopted.

This paper explores the feasibility of fully \textit{orthogonalizing} the far-field channels connecting multiple users to a common antenna array.
The rationale to pursue orthogonality is two-fold. Firstly, the trivialization of the precoder and receiver structures, rendering unnecessary the inversion of the channel matrix and
dramatically reducing the complexity when the numbers of antennas and users are large.
Secondly, the performance improvement that, in a variety of situations, can even lead to optimality.

An example that evidences the aforementioned advantages
is zero-forcing (ZF) uplink reception, whose achievable spectral efficiency is
\begin{align}
    \sum_k \log_2 \!\left(1+\frac{\SNR_k}{[\bG^{-1}]_{k,k}}\right)
\end{align}
where $\bG \equiv \bH^*\bH$ with $\bH \in\bbC^{N\times K}$ the channel matrix connecting an $N$-antenna array with $K$ single-antenna users, while $\SNR_k$ is a suitable signal-to-noise ratio for user $k$.
The sum-capacity 
of the corresponding orthogonal channel, in turn, would be determined by
\begin{align}
    \sum_k \log_2 \! \big(1+[\bG]_{k,k}\SNR_k \big).
\end{align}
Therefore, the power offset of ZF relative to the sum-capacity is \cite[Sec. 4.2]{foundations2018}
\begin{align}
    \bigg(\prod_k [\bG]_{k,k}[\bG^{-1}]_{k,k}\bigg)^{\!1/K}. \label{power_offset}
\end{align}
From Hadamard's inequality, the above satisfies 
\begin{align}
    \prod_k [\bG]_{k,k}[\bG^{-1}]_{k,k} & =  \frac{ \prod_k[\bG]_{k,k}}{\det\bG}\cdot \frac{\prod_k[\bG^{-1}]_{k,k}}{\det \bG^{-1}}  \ge 1,
\end{align}
with equality---indicative of optimality---if and only if the channel is orthogonal.
The benefits of such optimality are exemplified in Fig.~\ref{fig:power_offset}, which depicts the empirical cumulative distribution function (CDF) of \eqref{power_offset} for $N=8$ with $K=5$ users. For a baseline uniform linear array (ULA) with half-wavelength spacing, channel orthogonalization
yields a hefty $11.3$-dB power gain on average.

Certain array structures have been devised to render the channel more orthogonal.
A couple of these are also exemplified in Fig.~\ref{fig:power_offset}, namely
a sparse ULA stretching to $20$ wavelengths \cite{bjornson2026antenna}, and a minimum redundancy array (MRA) in which the number of redundant antenna spacings is minimized \cite{moffet1968minimum}. Although the power offset does shrink, these elaborate designs still incur respective offsets of $8.0$ and $4.3$ dB on average.
In the sequel, it will be shown that the offset can be eradicated entirely with eight antennas, as in Fig.~\ref{fig:graph}. 

The contributions of the paper are as follows.
\begin{itemize}
    \item For a broad class of arrays, a graph-theoretic framework is unveiled that, by mapping pairwise user channel correlations to graph edges, translates the continuous geometry problem of far-field channel orthogonalization into a discrete graph-covering problem on a complete graph. This framework is versatile, accommodating multipath channels and problem variants such as null steering. 
    \item Given arbitrarily positioned users, the minimum number of antennas required to achieve orthogonality is established for 1D, 2D, and 3D arrays. 
    A binary linear program is then formulated that identifies the most compact array---in the sense of the least mean width---attaining such orthogonality.
    \item A key tradeoff between exact and near orthogonality is uncovered: relaxing to near-orthogonality enables a reduction in antenna count at the expense of a penalty in array compactness.
    \item A method is devised to enforce antenna separation constraints (e.g., half-wavelength spacing) without sacrificing orthogonality or substantially expanding the array.
\end{itemize}

The manuscript is organized as follows. The problem is formulated in Sec.~\ref{sec:system_model} and 
the class of arrays being considered is described in Sec.~\ref{sec:arrays}. In Sec.~\ref{sec:fewer_antennas},
the graph-theoretic framework is set forth and
the minimum number of antennas required for exact orthogonality is established. Subsequently, in Sec.~\ref{sec:cover_optimization}, the most compact array attaining such orthogonality is identified. The versatility
of the graph-theoretic framework is discussed in Sec.~\ref{sec:versatility}.
The tradeoff among antenna count, array compactness, and the degree of orthogonality is the subject of Sec.~\ref{sec:near_orthogonality}. Then, Sec.~\ref{sec:minsep} describes the method to enforce
separation constraints without sacrificing orthogonality. Finally, Sec.~\ref{sec:conclusion} concludes the paper.

\pgfplotstableread[col sep=comma]{offset_avg.csv}\offsettable
\pgfplotstablegetelem{0}{avg_power_offset_dB}\of\offsettable
\edef\avgCompact{\pgfplotsretval}
\pgfplotstablegetelem{1}{avg_power_offset_dB}\of\offsettable
\edef\avgSparse{\pgfplotsretval}
\pgfplotstablegetelem{2}{avg_power_offset_dB}\of\offsettable
\edef\avgMRA{\pgfplotsretval}
\begin{figure*}[t]
    \centering
    \begin{tikzpicture}
    \begin{axis}[
        width=9cm, height=3cm,
        xlabel={Position [$\lambda$]},
        xmin=-1, xmax=21,
        ymin=-0.7, ymax=2.7,
        ytick={0,1,2},
        yticklabels={MRA, Sparse ULA, Baseline ULA},
        axis y line*=left,
        axis x line*=bottom,
        tick align=outside,
        very thick,
    ]
    \addplot[blue!70, only marks, mark=square*, mark size=1pt] coordinates {
        (0,2) (0.5,2) (1,2) (1.5,2) (2,2) (2.5,2) (3,2) (3.5,2)
    };
    \addplot[red!70, only marks, mark=square*, mark size=1pt] coordinates {
        (0,1) (2.857,1) (5.714,1) (8.571,1) (11.429,1) (14.286,1) (17.143,1) (20,1)
    };
    \addplot[orange!70, only marks, mark=square*, mark size=1pt] coordinates {
        (0,0) (4,0) (9,0) (9.5,0) (11,0) (12,0) (15.5,0) (19.5,0)
    };
    \end{axis}
    \end{tikzpicture}
    \begin{tikzpicture}
    \begin{axis}[
        width=7cm, height=3cm,
        xlabel={Power offset [dB]},
        ylabel={CDF},
        xmin=0, xmax=15,
        ymin=0, ymax=1,
        ytick={0,0.5,1.0},
        ylabel style={yshift=-3mm},
        legend pos=south east,
        legend cell align=left,
        legend style={font=\small, draw=none, fill=none},
        very thick,
    ]
    \addplot[blue!70]      table[x=offset_dB, y=compact_ula,   col sep=comma] {offset_cdf.csv};
    \addplot[red!70]     table[x=offset_dB, y=sparse_ula,   col sep=comma] {offset_cdf.csv};
    \addplot[orange!70] table[x=offset_dB, y=mra,   col sep=comma] {offset_cdf.csv};
    \addplot[blue!70, dotted, very thick, mark=none]   coordinates {(\avgCompact,0) (\avgCompact,1)};
    \addplot[red!70, dotted, very thick, mark=none]    coordinates {(\avgSparse,0)  (\avgSparse,1)};
    \addplot[orange!70, dotted, very thick, mark=none] coordinates {(\avgMRA,0)     (\avgMRA,1)};
    \end{axis}
    \end{tikzpicture}
    \caption{Left: array topologies for the baseline ULA, sparse ULA, and MRA.
    Right: CDF of the power offset for $K=5$ and $N=8$, obtained over a hundred thousand samples with arrival angles sampled uniformly from $-45^\circ$ to $45^\circ$ relative to broadside; the vertical dotted lines indicate the average values.}
    \label{fig:power_offset}
\end{figure*}
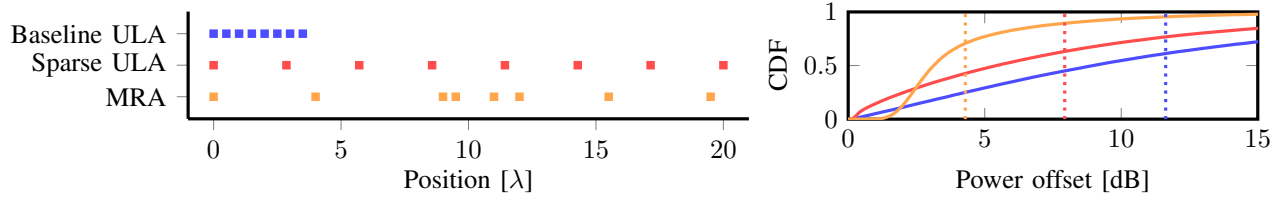

\section{Problem Formulation}
\label{sec:system_model}
\subsection{System Model}

As advanced, a connection between an $N$-antenna array and $K$ single-antenna users is considered. The coordinate system is chosen so that the array is in the vicinity of the origin.
The coordinate of the $n$th antenna, normalized by the wavelength, is $\bx_n\in U$, where $U$ is a subspace of $\bbR^3$ and $d=\dim U$. A volumetric array corresponds to $d=3$, while linear and planar arrays correspond to $d = 1$ and $d = 2$, respectively.

For any unit direction $\hat{\br}\in\bbR^3$, let us define the array steering vector $\ba(\hat{\br})\in\bbC^N$ whose $n$th entry is $e^{j 2\pi \langle \hat{\br}, \bx_n \rangle}$.
Denoting the coordinate of the $k$th user 
by $\br_k$, its direction is $\hat{\br}_k = \frac{\br_k}{\|\br_k\|}$ and the normalized far-field LOS channel between the array and the $k$th user is $\bh_k=\ba(\hat{\br}_k)$. The channel matrix is then
$
    \bH = [\bh_1\;\cdots\;\bh_K] .
$
   
Given $\bPi$ as the orthogonal projection matrix onto $U$, let us further define
\begin{align}
    \bp_k=\bPi\hat{\br}_k
\end{align}    
and
\begin{align}
    \bee_{k,\ell}=\bp_\ell-\bp_k\in U.
\end{align}
As $\bx_n\in U$, it holds that
$\langle\hat{\br}_k,\bx_n\rangle=\langle\bp_k,\bx_n\rangle$.
Thus, the channel depends on each user's direction only through $\bp_k$.

\subsection{Orthogonality Condition}

The objective in this paper is to render the channel orthogonal.
This amounts to $\bG = \bH^* \bH$ being diagonal and, under the specified normalization $\|\bh_k\|^2=N$, further to $\bG=N\bI$.
As the correlation between $\bh_k$ and $\bh_\ell$ is 
\begin{align}
    \frac{1}{N}[\bG]_{k,\ell}
    = \frac{1}{N}\sum_{\bx\in X} e^{j2\pi \langle \bee_{k,\ell}, \bx\rangle},
\end{align}
where $X = \{\bx_1,\ldots,\bx_N\} \subset U$,
the function $f_X:U \to\bbC$
\begin{align}
    f_X(\bee) = \frac{1}{|X|}\sum_{\bx\in X}e^{j2\pi \langle \bee, \bx\rangle}, \label{function_f}
\end{align}
allows recasting the orthogonality condition $\bG=N\bI$ into
\begin{align}
    f_X(\bee_{k,\ell}) = \begin{cases}
        0 &  k\ne \ell\\
        1 & k = \ell
    \end{cases}. \label{orthogonality_condition}
\end{align}

\section{The Zonotope Array} 
\label{sec:arrays}

This paper considers arrays of the form
\begin{align}
    X = X_1 + \cdots + X_M, \label{minkowski_array}
\end{align}
where $X_m\subset U$ for all $m$ and the summation is a Minkowski sum of sets.\footnote{Throughout the paper, the Minkowski sum double-counts identical elements. For instance, $\{0,1\}+\{0,1\} = \{0,1,1,2\}$.}
We shall term the above a \textit{Minkowski array},
for which the number of antennas satisfies
\begin{align}
    N = |X| = \prod_m |X_m|.
\end{align}
Any array is trivially a Minkowski array for $M=1$, hence \eqref{minkowski_array} entails no loss of generality. Its value rests on this result.

\begin{theorem}
\label{thm:minkowski_array}
For the Minkowski array in \eqref{minkowski_array}, it holds that
\begin{align}
    f_X(\bee) = \prod_{m} f_{X_m}(\bee). \label{decomposition}
\end{align}
\end{theorem}
\begin{proof}
See App. \ref{CL}.
\end{proof}

Intuitively, the fact that the orthogonality condition factorizes across the $X_m$ means that 
the necessary nulls can be distributed among the factors and handled one at a time.

\begin{figure}[t]
  \centering
  \begin{tikzpicture}[
      uvert/.style  = {circle, draw=black, fill=black!10, line width=0.5pt,
                       inner sep=0pt, minimum size=10pt, font=\small},
      ray/.style    = {black!40, line width=0.5pt},
      dome/.style   = {black!30, line width=0.4pt},
      face/.style   = {fill=black!12, fill opacity=0.55},
      gen/.style    = {line width=1pt, line cap=round}]
    \begin{scope}[x={(0.3632cm,0.0411cm)},
                  y={(-0.0973cm,0.1535cm)}, z={(0.0000cm,0.3408cm)}]
      \fill[blue!5]
        (1.88cm,0cm) arc[start angle=0, end angle=180, radius=1.88cm]
        -- plot[domain=-195:-15, samples=37] ({5*cos(\x)}, {5*sin(\x)}, 0) -- cycle;
      \coordinate (o) at (0,0,0);
      \foreach \k/\ux/\uy/\uz in {1/0.3130/0.9089/0.2756, 2/-0.2650/0.0760/0.9613,
                                  3/-0.4394/-0.3687/0.8192, 4/0.4465/-0.8397/0.3090,
                                  5/0.7637/0.5755/0.2924}
        \coordinate (w\k) at ({5*\ux}, {5*\uy}, {5*\uz});
      \fill[face] (-1.5152, 0.4000, -1.1559) -- (0.2187, -0.2585, -1.2235) -- (0.9286, -0.7625, -0.5082) -- (-0.8053, -0.1040, -0.4405) -- cycle;
      \fill[face] (-1.1078, 0.6518, 0.0000) -- (-1.5152, 0.4000, -1.1559) -- (-0.8053, -0.1040, -0.4405) -- (-0.6500, -0.0080, -0.0000) -- cycle;
      \fill[face] (-0.6500, -0.0080, 0.0000) -- (-0.8053, -0.1040, -0.4405) -- (0.9286, -0.7625, -0.5082) -- (1.1078, -0.6518, 0.0000) -- cycle;
      \draw[gen, red!70] (-1.1078, 0.6518, 0.0000) -- (-1.5152, 0.4000, -1.1559);
      \draw[gen, red!70] (-0.6500, -0.0080, 0.0000) -- (-0.8053, -0.1040, -0.4405);
      \draw[gen, red!70] (0.6500, 0.0080, 0.0000) -- (0.2187, -0.2585, -1.2235);
      \draw[gen, red!70] (1.1078, -0.6518, 0.0000) -- (0.9286, -0.7625, -0.5082);
      \draw[gen, blue!70] (-1.5152, 0.4000, -1.1559) -- (0.2187, -0.2585, -1.2235);
      \draw[gen, orange!70] (-1.5152, 0.4000, -1.1559) -- (-0.8053, -0.1040, -0.4405);
      \draw[gen, blue!70] (-0.8053, -0.1040, -0.4405) -- (0.9286, -0.7625, -0.5082);
      \draw[gen, orange!70] (0.2187, -0.2585, -1.2235) -- (0.9286, -0.7625, -0.5082);
      \fill[black] (-1.5152, 0.4000, -1.1559) circle (1.3pt);
      \fill[black] (-0.8053, -0.1040, -0.4405) circle (1.3pt);
      \fill[black] (0.2187, -0.2585, -1.2235) circle (1.3pt);
      \fill[black] (0.9286, -0.7625, -0.5082) circle (1.3pt);
      \fill[black!10, fill opacity=0.6]
        plot[domain=0:360, samples=73, smooth cycle] ({5*cos(\x)}, {5*sin(\x)}, 0);
      \draw[dome, black!18] plot[domain=-15:165, samples=37] ({5*cos(\x)}, {5*sin(\x)}, 0);
      \foreach \k in {1,5} {
        \draw[ray] (o) -- (w\k);
        \node[uvert] at (w\k) {\k};
      }
      \fill[blue!5, fill opacity=0.55]
        (1.88cm,0cm) arc[start angle=0, end angle=180, radius=1.88cm]
        -- plot[domain=-195:-15, samples=37] ({5*cos(\x)}, {5*sin(\x)}, 0) -- cycle;
      \fill[face] (-0.9286, 0.7625, 0.5082) -- (-1.1078, 0.6518, 0.0000) -- (-0.6500, -0.0080, -0.0000) -- (-0.2187, 0.2585, 1.2235) -- cycle;
      \fill[face] (-0.2187, 0.2585, 1.2235) -- (-0.6500, -0.0080, 0.0000) -- (1.1078, -0.6518, 0.0000) -- (1.5152, -0.4000, 1.1559) -- cycle;
      \draw[gen, red!70] (-0.9286, 0.7625, 0.5082) -- (-1.1078, 0.6518, 0.0000);
      \draw[gen, blue!70] (-0.9286, 0.7625, 0.5082) -- (0.8053, 0.1040, 0.4405);
      \draw[gen, orange!70] (-0.9286, 0.7625, 0.5082) -- (-0.2187, 0.2585, 1.2235);
      \draw[gen, red!70] (-0.2187, 0.2585, 1.2235) -- (-0.6500, -0.0080, 0.0000);
      \draw[gen, blue!70] (-0.2187, 0.2585, 1.2235) -- (1.5152, -0.4000, 1.1559);
      \draw[gen, red!70] (0.8053, 0.1040, 0.4405) -- (0.6500, 0.0080, 0.0000);
      \draw[gen, orange!70] (0.8053, 0.1040, 0.4405) -- (1.5152, -0.4000, 1.1559);
      \draw[gen, red!70] (1.5152, -0.4000, 1.1559) -- (1.1078, -0.6518, 0.0000);
      \fill[black] (-0.9286, 0.7625, 0.5082) circle (1.3pt);
      \fill[black] (-0.2187, 0.2585, 1.2235) circle (1.3pt);
      \fill[black] (0.8053, 0.1040, 0.4405) circle (1.3pt);
      \fill[black] (1.5152, -0.4000, 1.1559) circle (1.3pt);
      \foreach \k in {2,3,4}
        \draw[ray] (o) -- (w\k);
      \draw[dome] (1.88cm,0cm) arc[start angle=0, end angle=180, radius=1.88cm];
      \draw[dome] plot[domain=-195:-15, samples=37] ({5*cos(\x)}, {5*sin(\x)}, 0);
      \foreach \k in {2,3,4}
        \node[uvert] at (w\k) {\k};
    \end{scope}

    \begin{scope}[
        shift={(3.9,0.42)},
        vertex/.style={circle, draw=black, thick, fill=black!10, minimum size=12pt, inner sep=0pt},
        scale=1.5
      ]

        \newcommand{\offsetedge}[4]{
            \draw[#1] ($(#2)!#4!90:(#3)$) -- ($(#3)!#4!-90:(#2)$);
        }

        \def\offset{1.5pt} 
        \tikzset{edge/.style={line width=1.5pt}}

        \foreach \i in {1,...,5} {
            \coordinate (v\i) at ({90 + (\i-1)*72}:0.8cm);
        }


        \offsetedge{edge, red!70}{v1}{v2}{0pt}       
        \offsetedge{edge, red!70}{v2}{v3}{0pt}
        \offsetedge{edge, red!70}{v3}{v4}{0pt}
        \offsetedge{edge, red!70}{v4}{v1}{0pt}

        \offsetedge{edge, blue!70}{v1}{v3}{0pt}
        \offsetedge{edge, blue!70}{v3}{v5}{0pt}
        \offsetedge{edge, blue!70}{v5}{v2}{0pt}
        \offsetedge{edge, blue!70}{v1}{v2}{\offset}  

        \offsetedge{edge, orange!70}{v1}{v5}{0pt}
        \offsetedge{edge, orange!70}{v5}{v4}{0pt}
        \offsetedge{edge, orange!70}{v4}{v2}{0pt}
        \offsetedge{edge, orange!70}{v1}{v2}{-\offset} 

        \foreach \i in {1,...,5} {
            \node[vertex] at (v\i) {\i};
        }
    \end{scope}
  \end{tikzpicture}
  \caption{Left: five users around a zonotope array of $2^3=8$ antennas, with its edges parallel to the generators $\bt_1,\bt_2,\bt_3$ drawn in red, blue, and orange. Right: each generator nulls the user pairs along the 4-cycle of its color, and the three 4-cycles cover all ten pairs.}
  \label{fig:graph}
\end{figure}
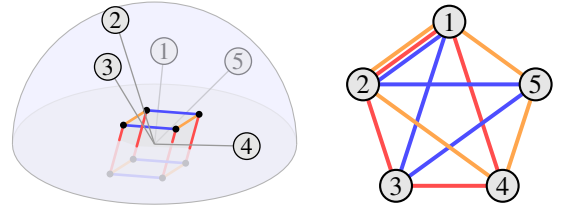

This paper considers Minkowski arrays satisfying
\begin{align}
    X_m = \frac{1}{2}\{-\bt_m,\bt_m\}, \label{zonotope_array}
\end{align}
which are referred to as \textit{zonotope arrays} because the convex hull of $X$ then forms a zonotope.
The above structure is no longer completely general, but this slight reduction in scope unlocks the analysis that follows.
(An example quantifying the impact of this restriction is provided in Sec.~\ref{sec:nonbinary_factors}.)
By its very structure, a zonotope array contains $2^M$ antennas. Fig.~\ref{fig:graph} (left) depicts one with $M=3$, whose antennas are the vertices of a parallelepiped.
Each $\bt_m$ will henceforth be referred to as a \textit{generator}.
A translation of $X$ multiplies $f_X$ by a unit-modulus factor and leaves the array size unchanged, hence an array remains essentially identical under translation; we accordingly use whichever representative is more convenient. 
For zonotope arrays, 
\begin{align}
    f_{X_m}(\bee) 
    &= \frac{1}{2}\sum_{\bx\in X_m}e^{j2\pi \langle \bee, \bx\rangle}\\
    &= \frac{1}{2}\big(e^{-j\pi \langle \bee, \bt_m\rangle}+e^{j\pi \langle \bee, \bt_m\rangle}\big)\\
    &=\cos(\pi \langle \bee, \bt_m\rangle). \label{cosine_factor}
\end{align}

\begin{remark}
The normalized beam pattern in direction $\hat{\br}$ with beamforming coefficients $\{w_\bx\}_{\bx\in X}$ is
\begin{align}
    \frac{1}{|X|}\sum_{\bx\in X} w_\bx e^{j2\pi \langle \hat{\br}, \bx\rangle}.
\end{align}
Although the beam pattern makes physical sense only over the unit sphere, $\|\hat{\br}\| = 1$, one can naturally extend it to $\bbR^3$. 
With that, one can interpret \eqref{function_f} as an extended beam pattern at $\bee$ with an all-ones beamformer, meaning a uniform illumination, by writing
\begin{align}
    f_X(\bee) = \frac{1}{|X|}\sum_{\bx\in X} 1\cdot e^{j2\pi \langle \bee, \bx\rangle}.
\end{align}
With the caveat that $\bee_{k,\ell}$ is not necessarily on the unit sphere, hence not physically meaningful, the orthogonality condition in \eqref{orthogonality_condition} can thus be interpreted as constructing nulls at $\bee_{k,\ell}$ for $k\ne \ell$ with an all-ones beamformer.
Then, Thm.~\ref{thm:minkowski_array} can be understood as a beam pattern decomposition  for the all-ones beamformer. This can be extended to the so-called separable beamformer, partly derived for the Minkowski sum of collinear ULAs resulting in arrays of arrays \cite{schelkunoff1943mathematical} or in fractal arrays \cite{puentebaliarda1996fractal, cohen2020sparse}, to two orthogonal ULAs resulting in a uniform planar array \cite{kummer1992basic}, and to three orthogonal ULAs resulting in cubic arrays \cite{wang2021beamforming}.
\end{remark}

\section{Minimum Antenna Count for Zonotope Arrays}
\label{sec:fewer_antennas}

For the zonotope array, \eqref{decomposition} and \eqref{cosine_factor} give
\begin{align}
    f_X(\bee_{k,\ell})
    = \prod_{m=1}^{M}\cos\!\big(\pi\langle \bee_{k,\ell},\bt_m\rangle\big).
\end{align}
Hence, users $k$ and $\ell$ are orthogonal if and only if 
\begin{align}
    \langle \bee_{k,\ell},\bt_m\rangle
    \equiv \tfrac12 \!\! \pmod 1 \label{generator_null_condition}
\end{align}
for some $m$. As a reflection of the interest in achieving orthogonality with the minimum number of antennas, we seek the smallest $M$ for which \eqref{generator_null_condition} holds for all $k\ne \ell$. It is shown next that this problem can be naturally
formulated as a graph covering problem. 
A graph also appears in \cite{mei2024movable} to model the minimum antenna spacing. Here, instead, the graph encodes which user pairs each generator renders orthogonal.

\subsection{Translation to Graph Theory Terminology}
\label{sec:translation_to_graph_theory}

To translate the problem into graph-theoretic terms, some terminology is needed, particularly in relation to graph covering \cite{knauer2016three}.
For a graph $G$, we let $V(G)$ and $E(G)$ denote its vertex and edge sets, respectively.

\begin{definition}
Let $H$ be a graph and let $\cG$ be a family of graphs on the vertex set $V(H)$.
A collection $G_1,\ldots,G_M\in\cG$ is called a
$\cG$-cover of $H$ if
\begin{align}
    E(H)=\bigcup_{m=1}^M E(G_m).
\end{align}
The number $M$ is the size of the cover. The
$\cG$-covering number of $H$ is the minimum size among all its $\cG$-covers, and a $\cG$-cover attaining it is called minimum.
\end{definition}

As an example with $\cG$ being a family of all 4-cycles, Fig.~\ref{fig:graph} (right) illustrates a cover of the complete graph on five vertices by three 4-cycles. 
One cannot cover ten edges with two 4-cycles, hence the $\cG$-covering number is three.

Identifying user $k$ with vertex $k$, carrying the vector
$\bp_k$, one can see that finding the minimum $M$ fulfilling \eqref{generator_null_condition} for all $k\ne \ell$ is equivalent to computing the $\cG$-covering number of $H$ with:
\begin{align*}
    H &= \text{the complete graph on $K$ vertices}\\
    \cG &= \text{a family of all graphs whose edge set is}\\
    &\qquad\qquad \big\{\{k,\ell\}:
    \langle \bee_{k,\ell},\bt\rangle
    \equiv \tfrac12 \!\! \pmod 1\big\}\\
    &\quad\,\, \text{for some $\bt\in U$}.
\end{align*}
Fig.~\ref{fig:graph} (left) shows an array realizing such a cover for $K=5$ users and $d=3$. As will be established in Cor.~\ref{cor:cover_family}, every 4-cycle can be realized by a single generator for $d=3$, provided the directions are in general position as defined next. Each of the array's three generators satisfies \eqref{generator_null_condition} for the user pairs along one 4-cycle, so $2^3=8$ antennas render all five users orthogonal.



\subsection{General Position Condition}
\label{sec:general_position}

What makes the computation of the $\cG$-covering number complicated is that $\cG$  seemingly depends on the directions $\hat{\br}_1,\ldots,\hat{\br}_K$.
The condition defined next, referred to as \textit{general position}, ensures that $\cG$ is independent of these directions.

\begin{definition}
\label{def:general_position}
The direction vectors $\hat{\br}_1,\ldots,\hat{\br}_K$ are in general position relative to $U$ if, for every forest $F$ on $\{1,2,\ldots,K\}$, there exists $\bt\in U$ satisfying
\begin{align}
    \langle\bee_{k,\ell},\bt\rangle\in\bbZ+\tfrac12 \text{ for all }\{k,\ell\}\in E(F)
    \label{general_position}
\end{align}
if and only if $|E(F)|\le d$.
\end{definition}

One can alternatively write \eqref{general_position} using a matrix. For a forest $F$ on $\{1,2,\ldots,K\}$, let $\bE_F\in\bbR^{3\times|E(F)|}$ be the matrix whose columns are labeled by the edges\footnote{This labeling implicitly identifies a finite set $A$ with $\{1,2,\ldots,|A|\}$. 
} $\{k,\ell\}\in E(F)$, with $k<\ell$, and whose $\{k,\ell\}$th column is $\bee_{k,\ell}$.
Thus, for $\bt\in U$, the $\{k,\ell\}$th entry of $\bE_F^\top\bt$ is $\langle\bee_{k,\ell},\bt\rangle$ and \eqref{general_position} can be written as
\begin{align}
    \bE_{F}^\top \bt
    \in \big(\bbZ+\tfrac{1}{2}\big)^{|E(F)|}.
\end{align}

Before proceeding, let us record a geometric consequence of general position.

\begin{lemma}
\label{lemma:affine_independence}
Suppose that the directions are in general position relative to $U$. Then, every set of at most $d+1$ of $\bp_1,\ldots,\bp_K$ is affinely independent.
\end{lemma}
\begin{proof}
See App.~\ref{app:affine_independence}.
\end{proof}




The following lemma shows that Def.~\ref{def:general_position} is unchanged if the half-integer values in \eqref{general_position} are prescribed arbitrarily for forests with at most $d$ edges.

\begin{lemma}
\label{lemma:forest_independence}
Suppose that the directions are in general position relative to $U$. Then, for every forest $F$ with at most $d$ edges, the columns of $\bE_F$ are linearly independent.
\end{lemma}
\begin{proof}
See App.~\ref{app:forest_independence}.
\end{proof}

Def.~\ref{def:general_position} would be of little value if it described a special configuration. The next result shows the opposite: the condition fails only for a null set of directions, holding with probability 1 for randomly located users; hence the term \textit{general} position.

\begin{theorem}
\label{thm:almost_sure_general_position}
If the joint distribution of the directions is absolutely continuous with respect to the product of surface measures on the unit sphere, then they are in general position relative to $U$ almost surely (a.s.).
\end{theorem}
\begin{proof}
See App.~\ref{app:almost_sure_general_position}.
\end{proof}
This covers not only directions drawn independently from distributions absolutely continuous with respect to surface measures, but also correlated directions.



\subsection{Characterization of the Family of Graphs}
\label{sec:family_of_graphs}
To compute the $\cG$-covering number of $H$, one first needs to characterize the family $\cG$.


\begin{lemma}
\label{lemma:no_odd_cycle}
For arbitrary directions, every nontrivial connected component of $G\in\cG$ is complete bipartite.
\end{lemma}
\begin{proof}
See App. \ref{Vitoria}.
\end{proof}

The following result characterizes $\cG$.

\begin{theorem}
\label{thm:cover_family}
Suppose that the directions are in general position relative to $U$. A graph on the $K$ vertices is in $\cG$ if and only if
\begin{enumerate}
    \item[(i)] every nontrivial connected component is complete bipartite;
    and
    \item[(ii)] some, equivalently every, spanning forest of the graph has at most $d$ edges.
\end{enumerate}
\end{theorem}
\begin{proof}
See App.~\ref{app:cover_family}.
\end{proof}

\begin{corollary}
\label{cor:cover_family}
Under general position relative to $U$, the family of graphs $\cG$ consists of the following:
\begin{enumerate}
    \item[(i)] for $d=1$, any graph with at most one edge;
    \item[(ii)] for $d=2$, any graph with at most two edges;
    \item[(iii)] for $d=3$, any forest with at most three edges, excluding paths with three edges, together with any cycle of length four.
\end{enumerate}
\end{corollary}

The empty graph cannot occur in a minimum cover of a nonempty graph $H$ and can be omitted from cover enumeration and optimization.
For $d=1$, the number of nonempty graphs in $\cG$ is $\binom{K}{2}$. For $d=2$, in turn, it is
\begin{align}
\binom{\binom{K}{2}}{2}+\binom{K}{2}.
\notag
\end{align}
For $d=3$, finally, it is
\begin{align*}
    \underbrace{3\binom{K}{4}}_{\text{$4$-cycles}}
    + \underbrace{\binom{\binom{K}{2}}{3}-\binom{K}{3}-12\binom{K}{4}}_{\text{three edges}}
    + \underbrace{\binom{\binom{K}{2}}{2}}_{\text{two edges}}
    + \underbrace{\binom{K}{2}}_{\text{one edge}}, 
\end{align*}
where, among all three-edge graphs, $\binom{K}{3}$ are cycles of length three and $12\binom{K}{4}$ are paths of length three. This number of nonempty graphs for $d=3$ evaluates to
\begin{align*}
    3+4+15+6 &= 28
    &&\text{for } K=4\\
    15+50+45+10 &= 120
    &&\text{for } K=5\\
    45+255+105+15 &= 420
    &&\text{for } K=6\\
    105+875+210+21 &= 1211
    &&\text{for } K=7\\
    210+2380+378+28 &= 2996
    &&\text{for } K=8\\
    378+5544+630+36 &= 6588
    &&\text{for } K=9\\
    630+11550+990+45 &= 13215
    &&\text{for } K=10.
\end{align*}
The generators of $G\in\cG$ can be computed explicitly, and from them the array geometry. Fix a spanning forest $F_G$ of $G$. A generator must produce every edge of $G$, hence the set of such generators is
\begin{align}
    \cT_G
    &=\big\{\bt\in U: \langle \bee_{k,\ell},\bt\rangle \in \bbZ+\tfrac12 \text{ for all } \{k,\ell\}\in E(G)\big\} \nonumber\\
    &=\big\{\bt\in U: \langle \bee_{k,\ell},\bt\rangle \in \bbZ+\tfrac12 \text{ for all } \{k,\ell\}\in E(F_G)\big\} \nonumber\\
    &=\big\{\bt\in U: \bE_{F_G}^\top \bt \in \big(\bbZ+\tfrac{1}{2}\big)^{|E(F_G)|}\big\}, \label{lattice}
\end{align}
where the second equality holds because every remaining edge of $G$ joins two vertices at odd distance in $F_G$. The graph generated by any $\bt\in\cT_G$ contains $G$, which suffices for covering. Under general position, the columns of $\bE_{F_G}$ are linearly independent (recall Lemma~\ref{lemma:forest_independence}), so each point of $(\bbZ+\tfrac12)^{|E(F_G)|}$ is attained on an affine subspace of dimension $d-|E(F_G)|$. Thus, $\cT_G$ is a disjoint union of such parallel subspaces, and a full-rank affine lattice when $|E(F_G)|=d$. The freedom in the choice of generators is exploited in Sec.~\ref{sec:cover_optimization}.

\subsection{Computation of the Covering Number}
\label{sec:computation_of_the_covering_number}

From Cor.~\ref{cor:cover_family}, the covering numbers for $d<3$ are
\begin{alignat}{2}
    &\tbinom{K}{2} &&\quad \text{for }d=1 \label{one_at_once}\\
    &\big\lceil\tbinom{K}{2}/2\big\rceil &&\quad \text{for }d=2. \label{two_at_once}
\end{alignat}
For $d=3$, each $G\in\cG$ has at most four edges, giving
\begin{align}
    \text{covering number}\ge \big\lceil\tbinom{K}{2}/4\big\rceil. \label{trivial_bound}
\end{align}
To determine the exact value, the packing and covering results for $4$-cycles in \cite{schonheim1975packing} are next applied. 

\begin{lemma}[{[\citen{schonheim1975packing}, Thms.~1 and~2]}]
\label{lemma:packing_number_cycles}
The maximum number of edge-disjoint 4-cycles that are subgraphs of the complete graph on $K$ vertices is
\begin{alignat}{2}
    &\big\lfloor \tfrac{K}{4} \big\lfloor  \tfrac{K-1}{2} \big\rfloor\big\rfloor-1 &&\quad \text{if }K \equiv 5, 7 \pmod 8 \label{packing_cycles}\\
    &\big\lfloor \tfrac{K}{4} \big\lfloor  \tfrac{K-1}{2} \big\rfloor\big\rfloor &&\quad \text{otherwise.} \nonumber
\end{alignat}
Moreover, these 4-cycles may be chosen so that the non-packed edges form
\begin{alignat*}{2}
    &\text{a perfect matching} &&\quad\text{if $K$ is even}\\
    & \text{an empty graph} &&\quad \text{if $K \equiv 1 \pmod 8$}\\
    &\text{a 3-cycle} &&\quad\text{if $K \equiv 3 \pmod 8$}\\
    &\text{two 3-cycles sharing a vertex} &&\quad\text{if $K \equiv 5 \pmod 8$}\\
    &\text{a 5-cycle} &&\quad\text{if $K \equiv 7 \pmod 8$}.
\end{alignat*}
\end{lemma}

\begin{lemma}[{[\citen{schonheim1975packing}, Thm.~3]}]
\label{lemma:covering_number_cycles}
For $K\ge4$, the minimum number of $4$-cycles whose union is the complete graph on $K$ vertices is
\begin{alignat}{2}
    &\big\lceil \tfrac{K}{4} \big\lceil  \tfrac{K-1}{2} \big\rceil\big\rceil + 1 &&\quad \text{if $K \equiv 3 \pmod 8$} \label{covering_cycles}\\
    &\big\lceil \tfrac{K}{4} \big\lceil  \tfrac{K-1}{2} \big\rceil\big\rceil &&\quad \text{otherwise}. \nonumber
\end{alignat}
\end{lemma}

Armed with these lemmas, the covering number for $d=3$ can be established.


\begin{theorem}
\label{thm:covering_number}
For $\cG$ in Cor. \ref{cor:cover_family} with $d=3$, the $\cG$-covering number of the complete graph $H$ is
\begin{alignat}{2}
        &\big\lfloor \tfrac{K}{4} \big\lfloor  \tfrac{K-1}{2} \big\rfloor\big\rfloor + \big\lceil \tfrac{K}{6}\big\rceil &&\quad \text{if $K$ is even} \label{covering_mixed}\\
        &\big\lfloor \tfrac{K}{4} \big\lfloor  \tfrac{K-1}{2} \big\rfloor\big\rfloor &&\quad \text{if } K \equiv 1 \pmod 8 \nonumber\\
        &\big\lfloor \tfrac{K}{4} \big\lfloor  \tfrac{K-1}{2} \big\rfloor\big\rfloor+2 &&\quad \text{if }K \equiv 3 \pmod 8 \nonumber\\
        &\big\lfloor \tfrac{K}{4} \big\lfloor  \tfrac{K-1}{2} \big\rfloor\big\rfloor+1 &&\quad \text{if }K \equiv 5 \pmod 8 \nonumber\\
        &\big\lfloor \tfrac{K}{4} \big\lfloor  \tfrac{K-1}{2} \big\rfloor\big\rfloor+1 &&\quad \text{if }K \equiv 7 \pmod 8. \nonumber
\end{alignat}
\end{theorem}
\begin{proof}
See App. \ref{app:covering_number}.
\end{proof}

Table~\ref{table:covering_number} lists the covering numbers \eqref{one_at_once}, \eqref{two_at_once}, and \eqref{covering_mixed} for $K = 2, 3, \ldots, 12 $.

\begin{table}
\setlength{\tabcolsep}{3pt}
    \centering
    \begin{threeparttable}
    \caption{
    Covering Numbers 
    }
    \label{table:covering_number}
    \begin{tabular}{cc|lllllllllll}
    \toprule
         & & \multicolumn{11}{c}{\#vertices ($K$)} \\
        $d$ & Eq. & 2 & 3 & 4 & 5 & 6 & 7 & 8 & 9 & 10 & 11 & 12\\
    \midrule
        1 & \eqref{one_at_once} & 1 & 3 & 6 & 10 & 15 & 21 & 28 & 36 & 45 & 55 & 66\\
        2 & \eqref{two_at_once} & 1 & 2 & 3 & 5 & 8 & 11 & 14 & 18 & 23 & 28 & 33 \\
        3 & \eqref{covering_mixed} & 1 & 2 & 2 & 3 & 4 & 6 & 8 & 9 & 12 & 15 & 17 \\
    \bottomrule
    \end{tabular}
    \end{threeparttable}
\end{table}

\begin{remark}
\label{remark:rotatable}
Suppose that the directions are in general position relative to $\bbR^3$.
For $K=4$, letting $U=\bbR^3$, Thm.~\ref{thm:covering_number} gives $M=2$, i.e., four antennas. The resulting array resides on the span of its two generators, hence is planar---with its plane dictated by the user directions. However, if $U$ were a plane specified beforehand and the directions were in general position relative to $U$, the construction by pairing two edges would require eight antennas. A planar array whose orientation is itself a design variable---a rotatable and movable array \cite{zhang2025roma}---thus attains the volumetric count.

The above argument nicely extends to $K\ge 6$. Let $U=\bbR^3$ and choose two generators, each satisfying the half-integer constraints of one of two edge-disjoint $4$-cycles. Then, take $V$ as the span of these generators and pair the remaining $\binom{K}{2}-8$ edges two at a time within $V$, yielding
\begin{align}
    M & = 2+ \bigg\lceil \frac{1}{2}\bigg(\binom{K}{2}-8\bigg) \bigg\rceil \\
    & = \bigg\lceil \frac{1}{2}\binom{K}{2} \bigg\rceil - 2, \label{num_generator_rotatable}
\end{align}
which is lower than \eqref{two_at_once} by two.
For $K=5$, no two $4$-cycles of the complete graph are edge-disjoint; two sharing an edge cover seven edges, leaving three for two more generators, so the saving reduces to one, i.e., $M=4$. Such pairing within $V$ is always possible by an appropriate choice of the generators (see App. \ref{app:pairing}).
\end{remark}

\begin{remark}
One can numerically compute the covering numbers, to reaffirm Thm. \ref{thm:covering_number}. For each $G\in\cG$, introduce a binary variable $y_G$ indicating whether $G$ is selected. Then, the problem of finding the covering number can be cast as the binary linear program
\begin{align}
    \min_{\{y_G\}}\quad & \sum_{G\in\cG} y_G \label{binary_programming_covering_number}\\
    \text{s.t.}\quad
    & \!\!\!\!\!\!\!\!\!\! \sum_{G:\{k,\ell\}\in E(G)} \!\!\!\!\!\!\!\!\!\!   y_G\ge 1 \text{ for all}\,\{k,\ell\}\in E(H) \nonumber\\
    &y_G\in\{0,1\} \text{ for all }G\in\cG.\nonumber
\end{align}
This formulation will come in handy in Sec. \ref{sec:cover_optimization}.
\end{remark}

\subsection{Extension to Minkowski Arrays with Non-Binary Factors}
\label{sec:nonbinary_factors}
The binary factors in \eqref{zonotope_array} may be relaxed to uniform linear factors \cite{zhu2023movable}
\begin{align}
    X_m=\{\zero,\bt_m,2\bt_m,\ldots,(b-1)\bt_m\} - \tfrac{b-1}{2}\bt_m,
    \label{ula_factor}
\end{align}
where $b> 1$ is an integer.
For $b=2$, the above reduces to \eqref{zonotope_array}. From
\begin{align}
    f_{X_m}(\bee) &= \frac{1}{|X_m|}\sum_{\bx\in X_m} e^{j2\pi \langle \bee, \bx\rangle}\\
    &= \frac{ \sin (b \pi \langle \bee, \bt_m\rangle)}{b \sin \pi \langle \bee, \bt_m\rangle},
\end{align}
which vanishes if and only if
\begin{align}
    \langle\bee,\bt\rangle
    \in \tfrac{1}{b}\bbZ\setminus\bbZ,
    \label{radix_null_condition}
\end{align} 
the relevant covering problem is 
\begin{align*}
    H &= \text{the complete graph on $K$ vertices}\\
    \cG_b &= \text{a family of all graphs whose edge set is}\\
    &\qquad\qquad \big\{\{k,\ell\}: 
    \langle \bee_{k,\ell},\bt\rangle
    \in \tfrac{1}{b}\bbZ\setminus\bbZ \big\}\\
    &\quad\,\, \text{for some $\bt\in U$}.
\end{align*}


\begin{lemma}
\label{lemma:multipartite}
For arbitrary directions, every nontrivial connected component
of $G\in\cG_b$ is complete multipartite with at most $b$ parts.
\end{lemma}

\begin{proof}
See App. \ref{ApH}.
\end{proof}

The notion of general position can be extended to an integer base $b\ge2$ by means of the
corresponding forest constraints.
\begin{definition}
\label{def:general_position_radix}
The directions $\hat{\br}_1,\ldots,\hat{\br}_K$ are in
general position for base $b$ relative to $U$ if, for every forest $F$
on $\{1,2,\ldots,K\}$,
\begin{enumerate}
    \item[(i)] the columns of $\bE_F$ are linearly independent when
    $|E(F)|\le d$; and
    \item[(ii)] no $\bt\in U$ satisfies \eqref{radix_null_condition}
    for all $\{k,\ell\}\in E(F)$ when $|E(F)|>d$.
\end{enumerate}
\end{definition}

This extension retains the desired properties of general position. The definition agrees with Def.~\ref{def:general_position} for $b=2$, and the assumption holds a.s. under the sampling conditions of Thm.~\ref{thm:almost_sure_general_position}.

\begin{theorem}
\label{thm:cover_family_radix}
Suppose that the directions are in general position for base $b$
relative to $U$. A graph $G$ on the $K$ vertices is in
$\cG_b$ if and only if
\begin{enumerate}
    \item[(i)] every nontrivial connected component is complete
    multipartite with at most $b$ parts; and
    \item[(ii)] some, equivalently every, spanning forest of the graph
    has at most $d$ edges.
\end{enumerate}
\end{theorem}

\begin{proof}
See App. \ref{ApI}.
\end{proof}

Two observations are in order. First, a triangle is complete tripartite, and for $K=3$ noncollinearity alone suffices for the following construction.

\begin{corollary}
\label{cor:three_antennas}
Provided that $d = \dim U \ge 2$ and $\bp_1,\bp_2,\bp_3$ are not collinear, there exists $\bt\in U$ satisfying
\begin{align}
    f_X(\bee_{1,2}) = f_X(\bee_{1,3}) = f_X(\bee_{2,3}) = 0,
\end{align}
where $X=\{\zero,\bt,2\bt\}$.
\end{corollary}
\begin{proof}
See App. \ref{ApJ}.
\end{proof}

Second, enlarging $b$ adds little. The graph associated with a single factor has a spanning forest with at most $d$ edges, hence it has at most $\binom{d+1}{2}$ edges regardless of how large $b$ is.

\begin{remark}
The construction in \cite{zhu2026multiuser} can be viewed as a special case of the non-binary Minkowski construction. The cross-linked planar array therein can be written, up to translation, as a Minkowski sum of factors
\begin{align}
    \{\zero,\bt_m,\ldots,(b_m-1)\bt_m\},
\end{align}
where each generator $\bt_m$ is constrained to lie along either the horizontal or vertical axis. For each user pair $\{k,\ell\}$, \cite{zhu2026multiuser} assigns one such factor and chooses its generator such that
\begin{align}
    \langle \bee_{k,\ell},\bt_m\rangle
    \in \tfrac{1}{b_m}\bbZ\setminus\bbZ,
\end{align}
which is precisely the condition in
\eqref{radix_null_condition}. Thus, the construction in \cite{zhu2026multiuser} corresponds
to covering the complete graph by singleton edges, one per factor. In contrast, Thm.~\ref{thm:cover_family_radix} allows a single factor to cover any graph in $\cG_{b_m}$ when its generator may point freely within $U$.
\end{remark}

\section{Cover Optimization for Array Compactness}
\label{sec:cover_optimization}

The covering number determines the smallest number of generators, and hence the smallest antenna count within the zonotope construction. A minimum cover is generally not unique, though, and different covers lead to different generator vectors and thus to different physical arrays.
This remaining degree of freedom can be leveraged to devise an array as compact as possible
while keeping the number of generators fixed at the covering number. Setting aside the obvious practical advantages of a compact form factor, small apertures are also preferred to reduce frequency selectivity \cite{do2024hybrid} and---if appropriate---to promote far-field propagation conditions \cite{do2026near}.

As a proxy of array compactness, we adopt the mean width of the array as an optimization objective. Since the array is a physical object in $\bbR^3$ regardless of $d$, letting $\bu$ be sampled from the uniform distribution over the unit sphere in $\bbR^3$, the mean width of $X$ is
\begin{align}
    \bbE\bigg[\max_{\bx,\bx'\in X} \langle\bu,\bx-\bx'\rangle\bigg] &= \sum_{m=1}^M \bbE\big[|\langle\bu,\bt_m\rangle|\big]\\
    &=\frac{1}{2} \sum_{m=1}^M \|\bt_m\|,
\end{align}
where $M$ is the $\cG$-covering number of the complete graph $H$.

\begin{remark}
Another meaningful measure of array compactness is the diameter of the smallest enclosing ball, which equals
\begin{align}
    2\max_{\bx\in X}\|\bx\| =\max_{\bs\in\{-1,1\}^{M}}
    \bigg\|\sum_{m=1}^M s_m\bt_m\bigg\|
    \label{exact_zonotope_aperture}
\end{align}
for the zonotope array in \eqref{zonotope_array}.
A direct evaluation of \eqref{exact_zonotope_aperture} entails testing $2^{M-1}$ sign vectors (opposite sign vectors yield the same norm). Polynomial-time methods avoid this exhaustive search \cite{allemand2001polynomial}; see \cite{do2026optimum} for a detailed description for $d=2$.

Alternatively, from
\begin{align}
    \|\bx\|=\max_{\|\bu\|=1}\langle\bu,\bx\rangle,
\end{align}
the diameter can also be expressed as
\begin{align}
    2\max_{\|\bu\|=1}
    \max_{\bx\in X}\langle\bu,\bx\rangle
    =\max_{\|\bu\|=1}
    \sum_{m=1}^M|\langle\bu,\bt_m\rangle|.
    \label{zonotope_maximum_width}
\end{align}
Replacing the maximum in \eqref{zonotope_maximum_width} by the average over the unit sphere in $\bbR^3$ \cite{do2026beamforming}, which is the mean width computed above, one obtains the sandwich bounds
\begin{align}
    \frac{1}{2}\sum_{m=1}^M\|\bt_m\|
    \le \max_{\|\bu\|=1}
    \sum_{m=1}^M|\langle\bu,\bt_m\rangle| \le\sum_{m=1}^M\|\bt_m\|
    \label{zonotope_aperture_bounds}
\end{align}
for every $d$, the upper bound being immediate from the Cauchy-Schwarz inequality. For $d=1$ and $2$, the lower bound can be tightened: since $\bt_1,\ldots,\bt_M\in U$, the maximum in \eqref{zonotope_maximum_width} is unchanged when $\bu$ is restricted to $U$, and averaging over the unit sphere in $U$ instead improves the factor $\frac{1}{2}$ to $1$ and $\frac{2}{\pi}$, respectively for $d=1$ and $d=2$. For $d=1$, the two bounds coincide, expressing the exact diameter.
\end{remark}

\subsection{Binary Linear Programming Formulation}

The optimization of the array compactness subject to the minimum cover can be formulated as 
\begin{align}
    \min_{G_1,\ldots,G_M\in\cG}\, & \min_{\bt_1\in\cT_{G_1}} \cdots \min_{\bt_M\in\cT_{G_M}} \sum_{m=1}^M \|\bt_m\| \\
    \text{s.t.}\quad & \bigcup_{m=1}^M E(G_m) = E(H). \nonumber
\end{align}
For fixed $G_1,\ldots,G_M$, the objective is a sum of terms each involving a single $\bt_m$, and each $\bt_m$ is constrained only by its own $G_m$. This separability allows carrying out the inner minimizations term by term, yielding
\begin{align}
\label{Yamal}
    \min_{G_1,\ldots,G_M\in\cG}\, & \sum_{m=1}^M \min_{\bt_m\in\cT_{G_m}} \|\bt_m\| \\
 \text{s.t.}\quad
    & \bigcup_{m=1}^M E(G_m) = E(H). \nonumber
\end{align}
Each term in \eqref{Yamal} depends on $G_m$ alone, hence it suffices to find, once for every $G\in\cG$, a shortest generator
\begin{align}
    \bt_G\in\argmin_{\bt\in\cT_G}\|\bt\|. \label{minimum_norm_solution}
\end{align}
For a nonempty $G$, writing $\bE_{F_G}^\top\bt=\bz+\frac12\one$ with $\bz\in\bbZ^{|E(F_G)|}$, the shortest $\bt\in U$ for a fixed $\bz$ is
\begin{align}
    \bt = \bE_{F_G} \big( \bE_{F_G}^\top\bE_{F_G} \big)^{-1}
    \left(\bz+\frac{1}{2}\one\right),
    \label{lattice_solution}
\end{align}
whereby \eqref{minimum_norm_solution} amounts to
\begin{align}
    \min_{\bz\in\bbZ^{|E(F_{G})|}} \left(\bz+\frac{1}{2}\one\right)^{\! \top}
    \big( \bE_{F_G}^\top\bE_{F_G} \big)^{-1} \left(\bz+\frac{1}{2}\one\right). \label{closest_vector_generator}
\end{align}
This is a closest-vector problem of dimension at most three, which can be solved exactly by sphere decoding. With $\|\bt_G\|$ precomputed for all $G\in\cG$, the problem in \eqref{Yamal} reduces to
\begin{align}
    \min_{G_1,\ldots,G_M\in\cG}\ & \sum_{m=1}^M \|\bt_{G_m}\| \\
    \text{s.t.}\quad & \bigcup_{m=1}^M E(G_m) = E(H). \nonumber
\end{align}
For each $G\in\cG$, introduce a binary variable $y_G$ indicating whether $G$ is selected. Then, the problem can be further recast as the binary linear program
\begin{align}
    \min_{\{y_G\}}\quad & \sum_{G\in\cG} \|\bt_G\| \, y_G \label{weighted_set_cover}\\
    \text{s.t.}\quad
    & \!\!\!\!\!\!\!\!\!\! \sum_{G:\{k,\ell\}\in E(G)} \!\!\!\!\!\!\!\!\!  y_G\ge 1 \text{ for all}\,\{k,\ell\}\in E(H)\nonumber\\
    &\sum_{G\in\cG} \, y_G=M \nonumber\\
    &y_G\in\{0,1\} \text{ for all }G\in\cG.\nonumber
\end{align}

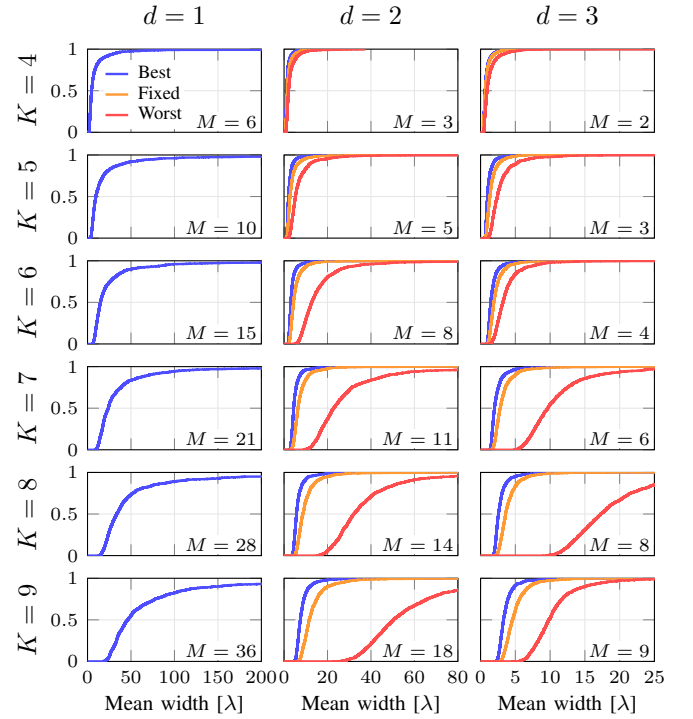
\begin{figure}[t]
    \centering
    \pgfplotsset{
        mean width axis/.style={
            width=2.3cm, height=1.1cm,
            scale only axis,
            xmin=0, ymin=0, ymax=1,
            ytick={0,0.5,1},
            xlabel={Mean width [$\lambda$]}, ylabel={},
            grid=major, major grid style={gray!20},
            xlabel style={yshift=2mm},
            tick label style={font=\scriptsize},
            label style={font=\footnotesize},
            title style={font=\normalsize},
            every axis plot/.append style={
                const plot, no marks, line width=0.9pt,
            },
            filter discard warning=false,
        },
        mean width best/.style={blue!70,solid},
        mean width fixed/.style={orange!80,solid},
        mean width worst/.style={red!70,solid},
    }
    \tikzset{
        generator count/.style={
            anchor=south east, font=\scriptsize,
            fill=white, inner sep=1pt,
            xshift=-0.5pt, yshift=0.5pt,
        },
    }
    \newcommand{\addMeanWidthCurves}[1]{%
        \addplot[mean width best] table[
            col sep=comma,x=mean_width_best,y=F] {#1};
        \addplot[mean width fixed] table[
            col sep=comma,x=mean_width_fixed,y=F] {#1};
        \addplot[mean width worst] table[
            col sep=comma,x=mean_width_worst,y=F] {#1};
    }
    \newcommand{\addMeanWidthUnique}[1]{%
        \addplot[mean width best] table[
            col sep=comma,x=mean_width_best,y=F] {#1};
    }
    \newcommand{\addGeneratorCount}[1]{%
        \node[generator count] at (rel axis cs:1,0) {$M=#1$};
    }
    \begin{tikzpicture}
        \begin{groupplot}[
            mean width axis,
            group style={
                group name=meanwidthplots,
                group size=3 by 6,
                horizontal sep=0.3cm,
                vertical sep=0.3cm,
                xlabels at=edge bottom,
                xticklabels at=edge bottom,
                ylabels at=edge left,
                yticklabels at=edge left,
            },
        ]
            \nextgroupplot[xmax=200,xtick={0,50,100,150,200},unbounded coords=discard,title={$d=1$},
                legend cell align=left,
                legend image code/.code={\draw[#1] (0cm,0cm) -- (0.3cm,0cm);},
                legend style={at={(rel axis cs:0.57,0.87)},anchor=north east,xshift=-1pt,draw=none,fill=white,font=\scriptsize,inner sep=0pt,row sep=-2.5pt,nodes={inner xsep=0.5pt},/tikz/every odd column/.append style={column sep=2.8pt}}]
                \addlegendimage{mean width best,no marks,line width=0.9pt}
                \addlegendentry{Best}
                \addlegendimage{mean width fixed,no marks,line width=0.9pt}
                \addlegendentry{Fixed}
                \addlegendimage{mean width worst,no marks,line width=0.9pt}
                \addlegendentry{Worst}
                \addMeanWidthUnique{mean_width_cdf_d1_K4.csv}
                \addGeneratorCount{6}

            \nextgroupplot[xmax=80,xtick={0,20,40,60,80},restrict x to domain=0:200,unbounded coords=discard,title={$d=2$}]
                \addMeanWidthCurves{mean_width_cdf_d2_K4.csv}
                \addGeneratorCount{3}

            \nextgroupplot[xmax=25,xtick={0,5,10,15,20,25},restrict x to domain=0:200,unbounded coords=discard,title={$d=3$}]
                \addMeanWidthCurves{mean_width_cdf_d3_K4.csv}
                \addGeneratorCount{2}

            \nextgroupplot[xmax=200,xtick={0,50,100,150,200},unbounded coords=discard]
                \addMeanWidthUnique{mean_width_cdf_d1_K5.csv}
                \addGeneratorCount{10}

            \nextgroupplot[xmax=80,xtick={0,20,40,60,80},restrict x to domain=0:200,unbounded coords=discard]
                \addMeanWidthCurves{mean_width_cdf_d2_K5.csv}
                \addGeneratorCount{5}

            \nextgroupplot[xmax=25,xtick={0,5,10,15,20,25},restrict x to domain=0:200,unbounded coords=discard]
                \addMeanWidthCurves{mean_width_cdf_d3_K5.csv}
                \addGeneratorCount{3}

            \nextgroupplot[xmax=200,xtick={0,50,100,150,200},unbounded coords=discard]
                \addMeanWidthUnique{mean_width_cdf_d1_K6.csv}
                \addGeneratorCount{15}

            \nextgroupplot[xmax=80,xtick={0,20,40,60,80},restrict x to domain=0:200,unbounded coords=discard]
                \addMeanWidthCurves{mean_width_cdf_d2_K6.csv}
                \addGeneratorCount{8}

            \nextgroupplot[xmax=25,xtick={0,5,10,15,20,25},restrict x to domain=0:200,unbounded coords=discard]
                \addMeanWidthCurves{mean_width_cdf_d3_K6.csv}
                \addGeneratorCount{4}

            \nextgroupplot[xmax=200,xtick={0,50,100,150,200},unbounded coords=discard]
                \addMeanWidthUnique{mean_width_cdf_d1_K7.csv}
                \addGeneratorCount{21}

            \nextgroupplot[xmax=80,xtick={0,20,40,60,80},restrict x to domain=0:200,unbounded coords=discard]
                \addMeanWidthCurves{mean_width_cdf_d2_K7.csv}
                \addGeneratorCount{11}

            \nextgroupplot[xmax=25,xtick={0,5,10,15,20,25},restrict x to domain=0:200,unbounded coords=discard]
                \addMeanWidthCurves{mean_width_cdf_d3_K7.csv}
                \addGeneratorCount{6}

            \nextgroupplot[xmax=200,xtick={0,50,100,150,200},unbounded coords=discard]
                \addMeanWidthUnique{mean_width_cdf_d1_K8.csv}
                \addGeneratorCount{28}

            \nextgroupplot[xmax=80,xtick={0,20,40,60,80},restrict x to domain=0:200,unbounded coords=discard]
                \addMeanWidthCurves{mean_width_cdf_d2_K8.csv}
                \addGeneratorCount{14}

            \nextgroupplot[xmax=25,xtick={0,5,10,15,20,25},restrict x to domain=0:200,unbounded coords=discard]
                \addMeanWidthCurves{mean_width_cdf_d3_K8.csv}
                \addGeneratorCount{8}

            \nextgroupplot[xmax=200,xtick={0,50,100,150,200},unbounded coords=discard]
                \addMeanWidthUnique{mean_width_cdf_d1_K9.csv}
                \addGeneratorCount{36}

            \nextgroupplot[xmax=80,xtick={0,20,40,60,80},restrict x to domain=0:200,unbounded coords=discard]
                \addMeanWidthCurves{mean_width_cdf_d2_K9.csv}
                \addGeneratorCount{18}

            \nextgroupplot[xmax=25,xtick={0,5,10,15,20,25},restrict x to domain=0:200,unbounded coords=discard]
                \addMeanWidthCurves{mean_width_cdf_d3_K9.csv}
                \addGeneratorCount{9}

        \end{groupplot}
        \node[font=\normalsize,rotate=90]
            at ([xshift=-0.8cm]meanwidthplots c1r1.west) {$K=4$};
        \node[font=\normalsize,rotate=90]
            at ([xshift=-0.8cm]meanwidthplots c1r2.west) {$K=5$};
        \node[font=\normalsize,rotate=90]
            at ([xshift=-0.8cm]meanwidthplots c1r3.west) {$K=6$};
        \node[font=\normalsize,rotate=90]
            at ([xshift=-0.8cm]meanwidthplots c1r4.west) {$K=7$};
        \node[font=\normalsize,rotate=90]
            at ([xshift=-0.8cm]meanwidthplots c1r5.west) {$K=8$};
        \node[font=\normalsize,rotate=90]
            at ([xshift=-0.8cm]meanwidthplots c1r6.west) {$K=9$};
    \end{tikzpicture}
    \caption{Empirical CDFs of the mean width for the best, fixed, and worst
        minimum covers, over 1000 realizations of $K$ directions uniformly distributed on the unit sphere in $\bbR^3$.
        Rows correspond to $K=4,\ldots,9$ and columns to $d=1,2,3$.
        Shown in each panel is its number of generators, $M$.
        For $d=1$, the minimum cover is unique and all three curves coincide. Each column uses a common horizontal range.}
    \label{fig:mean_width_cdf}
\end{figure}

Crucially, the optimization of the array compactness subject to the minimum cover cannot generally be solved by enumerating such minimum covers, as their number can be enormous; for $K=8$ and $K=10$, there are respectively 
9\,865\,742\,670 and 43\,957\,304\,088\,075
minimum covers (see App.~\ref{app:counting_covers} for the counting method). Thus, explicit enumeration is already prohibitive for those cases, whereas the above weighted set-covering formulation optimizes over them implicitly.

Also relevant is that
the cover returned by \eqref{weighted_set_cover} depends on the realized user directions through $\|\bt_G\|$. As an alternative, one can consider a single cover chosen without knowledge of those realizations, only of the generators computed from them via \eqref{closest_vector_generator}. Its expected mean width, $\frac{1}{2}\sum_{m=1}^M\bbE\big[\|\bt_{G_m}\|\big]$, is minimized by again solving \eqref{weighted_set_cover}, only with $\|\bt_G\|$ replaced by $\bbE\big[\|\bt_G\|\big]$ (with expectation over the distribution of user directions). For i.i.d. directions, relabeling the users leaves their joint distribution unchanged, hence $\bbE\big[\|\bt_G\|\big]$ depends only on the isomorphism class of $G$. From Cor.~\ref{cor:cover_family}, the nonempty graphs in $\cG$ fall into three such classes for $d=2$ and seven for $d=3$, hence only a handful of expectations need to be computed by Monte-Carlo runs. The result is termed \textit{fixed} cover,
and from it the generators---which do depend on the directions---can be computed
via \eqref{lattice}.

\subsection{Numerical Results}

Fig.~\ref{fig:mean_width_cdf} compares, for $d=1,2,3$ and various $K$, the mean widths obtained from minimum covers. 
For $d=1,2$, $U$ is a fixed line or plane whose orientation is immaterial because the user directions are isotropic.
For every candidate graph, the shortest generator is computed by sphere decoding of \eqref{closest_vector_generator}.

The \textit{best} curve is obtained by minimizing
\eqref{weighted_set_cover} while the \textit{worst} curve maximizes the same objective under the same constraints. Both problems are solved with the HiGHS optimizer through \texttt{SciPy}, after discarding the graphs that cannot appear in a minimum cover because the other $M-1$ graphs could not cover the remaining edges.
The \textit{fixed} curve uses the fixed cover for all realizations. For $d=2$ and $d=3$, optimizing for compactness pays off, and most of the gain over the worst cover is already achieved by the fixed cover.

\section{Versatility of the Graph Framework}
\label{sec:versatility}

\begin{figure}[t]
    \centering
    \begin{tikzpicture}[
        vertex/.style={
            circle, draw, fill=black!10,
            minimum size=12pt, inner sep=0pt
        }
    ]
        \def\Rp{1.2}
        \def\da{12}
        \def\dx{3.4}
        \def\dy{3.1}

        \node[font=\footnotesize] at (0,1.75) {Single path};
        \node[font=\footnotesize] at (\dx,1.75) {Two paths per user};
        \node[font=\footnotesize, rotate=90] at (-1.85,0) {Null steering};
        \node[font=\footnotesize, rotate=90] at (-1.85,-\dy) {Orthogonalization};

        \begin{scope}
            \foreach \i in {1,...,5} {
                \coordinate (a\i) at ({90 + 72 * (\i - 1)}:\Rp);
            }

            \foreach \i in {2,...,5} {
                \draw (a1) -- (a\i);
            }

            \foreach \i in {1,...,5} {
                \node[vertex] at (a\i) {\i};
            }
        \end{scope}

        \begin{scope}[yshift=-\dy cm]
            \foreach \i in {1,...,5} {
                \coordinate (d\i) at ({90 + 72 * (\i - 1)}:\Rp);
            }

            \foreach \i in {1,...,5} {
                \foreach \j in {1,...,5} {
                    \ifnum\i<\j
                        \draw (d\i) -- (d\j);
                    \fi
                }
            }

            \foreach \i in {1,...,5} {
                \node[vertex] at (d\i) {\i};
            }
        \end{scope}

        \begin{scope}[xshift=\dx cm]
            \foreach \g in {0,...,4} {
                \pgfmathsetmacro{\ca}{90 + 72*\g}
                \foreach \i in {0,1} {
                    \pgfmathsetmacro{\na}{\ca - \da + 2*\da*\i}
                    \coordinate (b\g\i)
                        at ({\Rp*cos(\na)},{\Rp*sin(\na)});
                }
            }

            \foreach \g in {1,...,4} {
                \foreach \i in {0,1} {
                    \foreach \j in {0,1} {
                        \draw (b0\i) -- (b\g\j);
                    }
                }
            }

            \foreach \g in {0,...,4} {
                \pgfmathtruncatemacro{\lab}{\g+1}
                \foreach \i in {0,1} {
                    \node[vertex] at (b\g\i) {\lab};
                }
            }
        \end{scope}

        \begin{scope}[xshift=\dx cm, yshift=-\dy cm]
            \foreach \g in {0,...,4} {
                \pgfmathsetmacro{\ca}{90 + 72*\g}
                \foreach \i in {0,1} {
                    \pgfmathsetmacro{\na}{\ca - \da + 2*\da*\i}
                    \coordinate (c\g\i)
                        at ({\Rp*cos(\na)},{\Rp*sin(\na)});
                }
            }

            \foreach \g in {0,...,4} {
                \foreach \h in {0,...,4} {
                    \ifnum\g<\h
                        \foreach \i in {0,1} {
                            \foreach \j in {0,1} {
                                \draw (c\g\i) -- (c\h\j);
                            }
                        }
                    \fi
                }
            }

            \foreach \g in {0,...,4} {
                \pgfmathtruncatemacro{\lab}{\g+1}
                \foreach \i in {0,1} {
                    \node[vertex] at (c\g\i) {\lab};
                }
            }
        \end{scope}
    \end{tikzpicture}

    \caption{Graph models for five users, with single-path (left) and two-path (right) propagation.
    Top: null steering toward user~1, with edges joining the vertices of user~1 to those of the interferers.
    Bottom: channel orthogonalization, with edges joining the vertices of every two distinct users; the single-path case is the complete graph of Sec.~\ref{sec:fewer_antennas}.
    Vertex labels identify users, each having one vertex per path.}
    \label{fig:graph_models}
\end{figure}
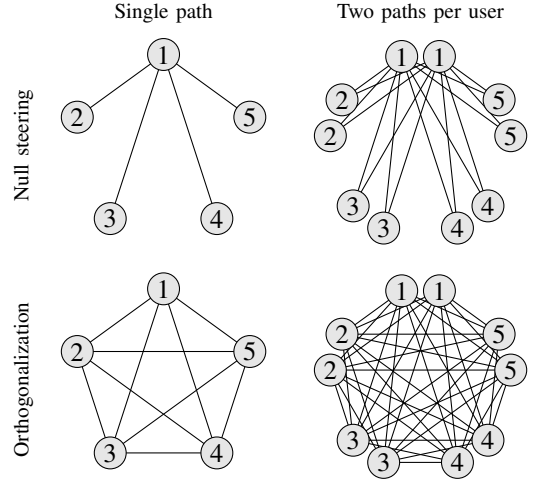

The graph-covering framework proposed in this paper is highly versatile. Many other problems on movable antenna design can be formulated as finding the $\cG$-cover of $H$ for
\begin{align*}
    H &= \text{some graph depending on the scenario of interest}\\
    \cG &= \text{a family of all graphs whose edge set is}\\
    &\qquad\qquad \big\{\{k,\ell\}: 
    \langle \bee_{k,\ell},\bt\rangle
    \equiv \tfrac12 \!\! \pmod 1\big\}  \cap E(H) \\
    &\quad\,\, \text{for some $\bt\in U$}.
\end{align*}
This section showcases three such examples, illustrated in Fig.~\ref{fig:graph_models} alongside the complete graph of Sec.~\ref{sec:fewer_antennas}.

\subsection{Null Steering}
\label{sec:null_steering}

Suppose that one intends to communicate with a single user, labeled 1, while minimizing the interference to/from $K-1$ directions, labeled $2,3,\ldots,K$.
Under maximum-ratio transmission/combining, the array optimization problem can be formulated as finding $X$ such that
\begin{align*}
    f_X(\bee_{k,\ell}) = 0 \text{ over }\{k,\ell\} \in \big\{\{1,2\}, \{1,3\}, \ldots, \{1,K\}\big\},
\end{align*}
which has been considered in \cite{zhu2023movable} for linear topologies.
This problem can be reformulated as a graph covering problem with $H$ being a star on $K$ vertices, which is a tree with $K-1$ leaves (Fig.~\ref{fig:graph_models}, top left). The covering number is
\begin{alignat}{2}
    &K-1 &&\quad \text{for }d=1, \\
    &\big\lceil\tfrac{K-1}{2}\big\rceil &&\quad \text{for }d=2,\\
    &\big\lceil\tfrac{K-1}{3}\big\rceil &&\quad \text{for }d=3.
\end{alignat}
For $d=1$, this recovers \cite[Thm. 1]{zhu2023movable}, where an $N$-antenna linear array is shown to be capable of nulling $K-1$ directions whenever $K-1$ does not exceed the number of prime factors of $N$ counted with multiplicity; the construction therein is a Minkowski array whose factors are ULAs, producing one null per factor. For $d=2$, the case $K=3$ is studied in \cite{mochaourab2011beamforming}.

A closely related null-steering problem is considered in \cite{ismail1991null}, where the antenna positions within a linear array are perturbed to place nulls in prescribed directions. The design, however, relies on a first-order approximation valid only for small position perturbations.

\subsection{Null Steering Under Multipath Propagation}

Suppose now that the channel undergoes multipath propagation. Let $L_k$ be the number of paths of user $k$, with directions $\hat{\br}_{k,p}$ and path gains $g_{k,p}\in\bbC$ for $p=1,\ldots,L_k$, so that
\begin{align}
    \bh_k = \sum_{p=1}^{L_k} g_{k,p}\, \ba(\hat{\br}_{k,p}),
    \label{multipath_channel}
\end{align}
with $\ba$ the steering vector (see Sec.~\ref{sec:system_model}). As in Sec.~\ref{sec:system_model}, define $\bp_{k,p}=\bPi\hat{\br}_{k,p}$ and the path differences
\begin{align}
\bee_{(k,p),(\ell,q)}=\bp_{\ell,q}-\bp_{k,p} .
\end{align}
The pairwise user correlations then expand over path pairs as
\begin{align}
    \frac{\langle \bh_k,\bh_\ell\rangle}{N}
    = \sum_{p=1}^{L_k}\sum_{q=1}^{L_\ell} \overline{g_{k,p}}\,g_{\ell,q}\, f_X\big(\bee_{(k,p),(\ell,q)}\big).
    \label{multipath_inner}
\end{align}
Under maximum-ratio transmission/combining, the interference from user $k$ is determined by $\langle\bh_1,\bh_k\rangle$, so a sufficient condition for nulling interferer $k$ is to null every path pair between user $1$ and user $k$, i.e.,
\begin{align}
    f_X\big(\bee_{(1,p),(k,q)}\big) = 0
\end{align}
for $p=1,\ldots,L_1$, $q=1,\ldots,L_k$, and $k=2,\ldots,K$.
Thus $H$ is the complete bipartite graph between the $L_1$ desired-user paths and the $\sum_{k=2}^K L_k$ interferer paths; Fig.~\ref{fig:graph_models} (top right) illustrates the case $K=5$ and $L_k=2$. When every user has a single path, this graph reduces to the star.

\subsection{Channel Orthogonalization Under Multipath Propagation}
\label{sec:multipath}

Under the multipath channel in \eqref{multipath_channel}, it is evidenced by \eqref{multipath_inner} that a sufficient condition for channel orthogonality is
\begin{align}
    f_X\big(\bee_{(k,p),(\ell,q)}\big) = 0
    \label{multipath_condition}
\end{align}
for all $k\ne\ell$, $p=1,\ldots,L_k$, and $q=1,\ldots,L_\ell$. This makes $\bG$ diagonal, although its diagonal entries need not equal $N$ under multipath propagation.


The condition in \eqref{multipath_condition} is a covering condition, now on a larger graph. Assigning a vertex to each of the $\sum_k L_k$ paths and grouping the $L_k$ vertices of user $k$ into a part, the pairs in \eqref{multipath_condition} are exactly the edges joining distinct parts, i.e., the complete multipartite graph (Fig.~\ref{fig:graph_models}, bottom right). The single-path case $L_k=1$ collapses each part to a vertex and yields the complete graph of Sec.~\ref{sec:fewer_antennas} (Fig.~\ref{fig:graph_models}, bottom left).

\section{Exact vs Near Orthogonality}
\label{sec:near_orthogonality}

This paper has thus far considered the problem of constructing an \textit{exactly} orthogonal channel. Remarkably, there is a major difference between \textit{exact} and \textit{near} orthogonality. It is shown in \cite{leshem2021interference} that, for any $\epsilon>0$ and almost all directions, a single generator (i.e., $X = \frac{1}{2}\{-\bt,\bt\}$) and therefore only two antennas can satisfy
\begin{align}
    |f_X(\bee_{k,\ell})| \le \epsilon \text{ over }\{k,\ell\} \in \big\{\{1,2\}, \{1,3\}, \ldots, \{1,K\}\big\}, \label{near_null}
\end{align}
which is a relaxed version of the null-steering problem formulated in Sec.~\ref{sec:null_steering}. Since $f_X(\bee_{1,k})=\cos(\pi\langle\bee_{1,k},\bt\rangle)$ by \eqref{cosine_factor}, the above condition only requires each $\langle\bee_{1,k},\bt\rangle$ to lie within $\delta=\frac{1}{\pi}\arcsin\epsilon$ of a half-integer.
Let $\bt=s\bu$ move along a fixed direction $\bu\in U$, such that $\langle\bee_{1,k},\bt\rangle=s\langle\bee_{1,k},\bu\rangle$. Provided that the numbers $\langle\bee_{1,2},\bu\rangle,\ldots,\langle\bee_{1,K},\bu\rangle$ are linearly independent over the rationals, which holds for almost all user directions, Kronecker's theorem \cite[p.~507]{gonek2016kronecker} guarantees, for every $\delta>0$, a real $s$ for which every $s\langle\bee_{1,k},\bu\rangle$ lies within $\delta$ of $\bbZ+\frac12$. A single generator thus suffices for any number of interferers, whereas exact nulls require $\lceil \frac{K-1}{d}\rceil$ generators.

\pgfplotstableread[col sep=comma]{near_null_d1_K3_cdf.csv}\nearnulltable
\pgfplotstablegetelem{400}{mean_width_near_eps0p1}\of\nearnulltable
\edef\nearLabelA{\pgfplotsretval}
\pgfplotstablegetelem{400}{mean_width_near_eps0p01}\of\nearnulltable
\edef\nearLabelB{\pgfplotsretval}
\pgfplotstablegetelem{400}{mean_width_near_eps0p001}\of\nearnulltable
\edef\nearLabelC{\pgfplotsretval}
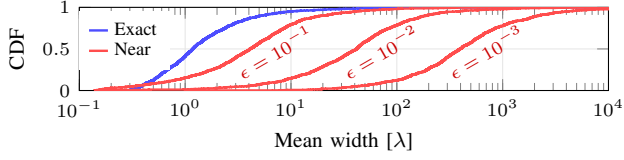
\begin{figure}[t]
    \centering
    \begin{tikzpicture}[
        epsilon label/.style={anchor=west, rotate=30, font=\scriptsize, text=red!80!black, inner sep=1pt},
    ]
    \begin{semilogxaxis}[
        width=7cm, height=1.1cm,
        scale only axis,
        xlabel={Mean width [$\lambda$]},
        ylabel={CDF},
        xlabel style={yshift=1mm},
        ylabel style={yshift=-4mm},
        xmin=0.1, xmax=10000,
        ymin=0, ymax=1,
        ytick={0,0.5,1},
        grid=major, major grid style={gray!20},
        legend columns=1,
        legend cell align=left,
        legend image code/.code={\draw[#1] (0cm,0cm) -- (0.25cm,0cm);},
        legend style={at={(0,0.60)},anchor=west, font=\scriptsize, draw=none, fill=white, row sep=-2.5pt},
        tick label style={font=\scriptsize},
        label style={font=\footnotesize},
        every axis plot/.append style={const plot, no marks, line width=0.9pt},
        unbounded coords=discard, filter discard warning=false,
    ]
    \addplot[blue!70] table[x=mean_width_exact, y=F] {\nearnulltable};
    \addlegendentry{Exact}
    \addplot[red!70] table[x=mean_width_near_eps0p1, y=F] {\nearnulltable};
    \addlegendentry{Near}
    \addplot[red!70, forget plot] table[x=mean_width_near_eps0p01, y=F] {\nearnulltable};
    \addplot[red!70, forget plot] table[x=mean_width_near_eps0p001, y=F] {\nearnulltable};
    \node[epsilon label] at (axis cs:\nearLabelA,0.2) {$\epsilon=10^{-1}$};
    \node[epsilon label] at (axis cs:\nearLabelB,0.2) {$\epsilon=10^{-2}$};
    \node[epsilon label] at (axis cs:\nearLabelC,0.2) {$\epsilon=10^{-3}$};
    \end{semilogxaxis}
    \end{tikzpicture}
    \caption{Empirical CDFs of the mean width for null steering with $d=1$ and $K=3$, over 1000 realizations of $K$ directions uniformly distributed on the unit sphere in $\bbR^3$: exact nulls with two generators vs a single generator satisfying \eqref{near_null}.}
    \label{fig:near_null}
\end{figure}

The price of this reduced antenna count with near orthogonality is a much less compact array,
as Fig.~\ref{fig:near_null} illustrates for $d=1$ and $K=3$. The exactly orthogonal design entails one generator per interferer, i.e., $M=2$ generators and four antennas, which is the unique minimum cover for $d=1$. The near-orthogonal design is the shortest single generator satisfying \eqref{near_null}, which we find exactly rather than over a grid of spacings as done in \cite{leshem2021interference}, to ensure that the comparison is not an artifact of a suboptimal search. Writing $\bt=s\bu$ for a unit vector $\bu$ spanning $U$, each condition in \eqref{near_null} confines $s$ to a union of intervals; the smallest $s>0$ meeting all of them is the left endpoint of one of these intervals, whereby checking the left endpoints in increasing order finds it.

The above evinces the existence of a tradeoff among antenna count, array size, and the level of orthogonality, which warrants further research.

\section{Minimum Antenna Separation}
\label{sec:minsep}

In the movable antenna literature, it is customary to keep the antennas at least half a wavelength apart to limit mutual coupling. Indeed, when the goal is to orthogonalize the channel, mutual coupling cannot be beneficial---its main potential is the possibility of superdirectivity in certain directions \cite{pizzo2025mutual,zhu2026mutual}.
The arrays of Sec.~\ref{sec:cover_optimization} need not respect the half-wavelength spacing. However, as $\kappa\cT_G\subset\cT_G$ for every odd integer $\kappa$, each generator $\bt_m$ can be replaced by an odd multiple $\kappa_m\bt_m$ without affecting the cover. 
This freedom always suffices because:
\begin{enumerate}
    \item The odd integers can be chosen successively such that the antennas lie at distinct positions. Indeed, the $m$th factor $\frac12\{-\kappa_m\bt_m,\kappa_m\bt_m\}$ places two translations of the antennas of the preceding factors, which overlap only if $\kappa_m\bt_m$ is the difference of two of these antennas. This excludes finitely many values of $\kappa_m$.
    \item Multiplying every $\kappa_m$ by a common odd integer keeps them odd and scales every separation by the same factor; a sufficiently large common odd integer then does it.
\end{enumerate}

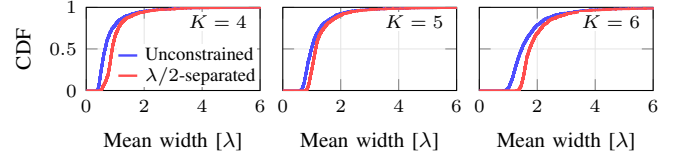
\begin{figure}[t]
    \centering
    \tikzset{
        user count/.style={
            anchor=south east, font=\scriptsize,
            fill=white, inner sep=1pt,
            xshift=-0.5pt, yshift=0.5pt,
        },
    }
    \begin{tikzpicture}
        \begin{groupplot}[
            group style={
                group size=3 by 1,
                horizontal sep=0.3cm,
                yticklabels at=edge left,
                ylabels at=edge left,
            },
            width=2.3cm, height=1.1cm,
            scale only axis,
            xmin=0, xmax=6, xtick={0,2,4,6},
            ymin=0, ymax=1, ytick={0,0.5,1},
            xlabel={Mean width [$\lambda$]}, ylabel={CDF},
            grid=major, major grid style={gray!20},
            xlabel style={yshift=1mm},
            ylabel style={yshift=-4mm},
            tick label style={font=\scriptsize},
            label style={font=\footnotesize},
            every axis plot/.append style={const plot, no marks, line width=0.9pt},
            filter discard warning=false,
        ]
            \nextgroupplot[
                legend cell align=left,
                legend image code/.code={\draw[#1] (0cm,0cm) -- (0.25cm,0cm);},
                legend style={at={(rel axis cs:1,0)},anchor=south east,xshift=-1pt,yshift=1pt,draw=none,fill=white,font=\scriptsize,inner sep=0pt,row sep=-0.9pt,nodes={inner xsep=0.5pt,inner ysep=1pt},/tikz/every odd column/.append style={column sep=2pt}}]
                \addplot[blue!70] table[col sep=comma, x=mean_width_unconstrained, y=F] {min_separation_cdf_d3_K4.csv};
                \addlegendentry{Unconstrained}
                \addplot[red!70] table[col sep=comma, x=mean_width_separated, y=F] {min_separation_cdf_d3_K4.csv};
                \addlegendentry{$\lambda/2$-separated}
                \node[user count] at (rel axis cs:0.95,0.7) {$K=4$};
            \nextgroupplot
                \addplot[blue!70] table[col sep=comma, x=mean_width_unconstrained, y=F] {min_separation_cdf_d3_K5.csv};
                \addplot[red!70] table[col sep=comma, x=mean_width_separated, y=F] {min_separation_cdf_d3_K5.csv};
                \node[user count] at (rel axis cs:0.95,0.7) {$K=5$};
            \nextgroupplot
                \addplot[blue!70] table[col sep=comma, x=mean_width_unconstrained, y=F] {min_separation_cdf_d3_K6.csv};
                \addplot[red!70] table[col sep=comma, x=mean_width_separated, y=F] {min_separation_cdf_d3_K6.csv};
                \node[user count] at (rel axis cs:0.95,0.7) {$K=6$};
        \end{groupplot}
    \end{tikzpicture}
    \caption{Empirical CDFs of the mean width for $d=3$ over the 1000 realizations of $K$ directions uniformly distributed on the unit sphere in $\bbR^3$.}
    \label{fig:min_separation}
\end{figure}

The above procedure is sure to return odd integers $\{\kappa_m\}$ such that $\{\kappa_m \bt_m\}$ generate an array with at least half-wavelength spacing. This does increase the mean width, but such excess is expected to be slight as corroborated 
by Fig.~\ref{fig:min_separation}, where
an exact search (see App. \ref{app:minimum_separation} for details) over $\{\kappa_m\}$ is conducted to construct an array with minimum mean width while abiding by the minimum separation constraint. The cost in mean width is indeed very modest.
And, the restriction to odd multiples leaves room for further optimization, e.g., over other lattice points of $\cT_G$.







\section{Conclusion}
\label{sec:conclusion}

The possibility of repositioning the antennas that constitute an array opens the door to optimizing the channel itself, rather than optimizing the precoders and receivers. 
This paper has pursued a specific optimization of the channel, namely orthogonalization, which
renders the precoders and receivers trivial and can improve the performance heftily.
This orthogonalization has been tackled within a graph-theoretic framework, whose versatility invites extensions beyond the ones already addressed in the paper.

One of the insights emanating from the results is that, the higher the array's geometric dimensionality, the fewer required antennas for a given number of users.
For 4 users, for instance, a linear array (recall Table \ref{table:covering_number}) requires $2^6=64$ antennas, but a planar array necessitates only $2^3=8$ antennas, and a volumetric array as few as $2^2=4$ antennas.
Another relevant insight is that a cover chosen from the distribution of user directions alone, with generators computed for the actual directions, already yields highly compact arrays (while the strictly most compact one follows from a binary linear program). 
Yet another insight is that minimum antenna spacings can be imposed straightforwardly, without compromising the orthogonalization.

Perhaps the most striking observation is that, if near-orthogonality suffices, then far fewer antennas can get the job done---but those antennas might have to be far apart. A relevant avenue for subsequent work would be to explore the tradeoff that arises among 
antenna count, array size, and the level of orthogonality.


\appendices

\section{}
\label{CL}
The decomposition in Thm. \ref{thm:minkowski_array} directly follows from the definition of the Minkowski sum:
\begin{align}
    \frac{1}{|X|}\sum_{\bx\in X}e^{j2\pi \langle \bee, \bx\rangle} & = \frac{1}{|X
    |}\sum_{\bx_1\in X_1}\cdots \sum_{\bx_M\in X_M} e^{j2\pi \langle \bee, \sum_m \bx_m\rangle} \notag \\
    & = \frac{1}{|X
    |}\sum_{\bx_1\in X_1}\cdots \sum_{\bx_M\in X_M} \prod_m  e^{j2\pi \langle \bee, \bx_m\rangle} \notag \\
    & = \prod_{m} \frac{1}{|X_m|}\sum_{\bx_m\in X_m} \!\! e^{j2\pi \langle \bee, \bx_m\rangle}\\
    & = \prod_{m} f_{X_m}(\bee).
\end{align}

\section{}
\label{app:affine_independence}

Let the directions be in general position relative to $U$, and consider $s\le d+1$ of the vectors $\bp_1,\ldots,\bp_K$, relabeled as $\bp_1,\ldots,\bp_s$. It is enough to show that $\bee_{1,2},\ldots,\bee_{1,s}$ are linearly independent, and the case $s=1$ is immediate.
For each $\ell=2,\ldots,s$, the star on the vertices $1,\ldots,s$ centered at $\ell$ has $s-1\le d$ edges, so doubling a solution of \eqref{general_position} for this star gives $\bv_\ell\in U$ such that $\langle\bee_{k,\ell},\bv_\ell\rangle$ is an odd integer for every $k\ne\ell$. Now, consider the integer matrix
\begin{align*}
    \begin{bmatrix}\bee_{1,2} & \cdots & \bee_{1,s}\end{bmatrix}^\top
    \begin{bmatrix}\bv_2 & \cdots & \bv_s\end{bmatrix}.
\end{align*}
From $\bee_{1,k} = \bee_{1,\ell}-\bee_{k,\ell}$, its $(k,\ell)$th entry is 
\begin{align}
    \langle\bee_{1,k},\bv_\ell\rangle = \langle\bee_{1,\ell},\bv_\ell\rangle-\langle\bee_{k,\ell},\bv_\ell\rangle. 
\end{align}
By construction, the first term is an odd integer. The second term, however, is zero when $k=\ell$ and odd otherwise.
Thus, the integer matrix is the identity matrix modulo 2, so its determinant is odd and hence nonzero. As the rank of a matrix product cannot exceed that of either factor, the vectors $\{\bee_{1,2},\ldots,\bee_{1,s}\}$ are linearly independent.

\section{}
\label{app:forest_independence}

Let the directions be in general position relative to $U$, and suppose that Lemma~\ref{lemma:forest_independence} fails. Among the forests with at most $d$ edges whose edge differences are linearly dependent, let $F$ be one with the largest number of edges, say $r\le d$. The edge differences of $F$ span a subspace of dimension at most $r-1$, which contains $\bee_{k,\ell}$ whenever $k$ and $\ell$ lie in the same tree of $F$. On the other hand, any $\min(K,d+1)$ of $\bp_1,\ldots,\bp_K$ are affinely independent by Lemma~\ref{lemma:affine_independence}, so the differences $\bee_{k,\ell}$ over all pairs span a subspace of dimension $\min(K-1,d)\ge r$; here $r\le K-1$ because $F$ is a forest. Hence some $\bee_{k,\ell}$ lies outside the span of the edge differences of $F$, with $k$ and $\ell$ in different trees of $F$. Adding $\{k,\ell\}$ to $F$ gives a forest with $r+1$ edges whose edge differences are still linearly dependent.

If $r<d$, this contradicts the maximality of $r$. If $r=d$, general position supplies $\bt_0\in U$ satisfying \eqref{general_position} for $F$. Let $\bu\in U$ be the component of $\bee_{k,\ell}$ orthogonal to the edge differences of $F$, so that $\langle\bee_{k,\ell},\bu\rangle\ne0$. Along $\bt_0+s\bu$, the constraints on $F$ remain satisfied, while $\langle\bee_{k,\ell},\bt_0+s\bu\rangle$ takes every real value, in particular $\frac{1}{2}$. The forest with $d+1$ edges then admits a solution of \eqref{general_position}, contradicting general position.

\section{}
\label{app:almost_sure_general_position}

As there are finitely many forests, it suffices to fix a forest $F$ and show that, a.s.,
\begin{enumerate}
    \item[(a)] if $|E(F)|\le d$, the columns of $\bE_F$ are linearly independent, so that \eqref{general_position} has a solution $\bt\in U$;
    \item[(b)] if $|E(F)|>d$, \eqref{general_position} has no solution $\bt\in U$.
\end{enumerate}
For (b), we may assume $|E(F)|=d+1$, since every larger forest contains a forest of this size.

Both parts above rest on the following result.

\begin{lemma}
\label{lemma:circle}
Under the assumption of Thm.~\ref{thm:almost_sure_general_position}, fix a user $\ell$, and let a nonzero $\bu\in U$ and a value be determined by the directions of the other users. Then $\langle\bp_\ell,\bu\rangle$ equals this value with probability zero.
\end{lemma}
\begin{proof}
Since $\bu\in U$, we have $\langle\bp_\ell,\bu\rangle=\langle\hat{\br}_\ell,\bu\rangle$. For fixed directions of the other users, the unit vectors $\hat{\br}_\ell$ attaining the value lie on a plane, forming a circle, a single point, or the empty set, of surface measure zero. Integrating over the other directions, the event has product surface measure zero, and hence probability zero by absolute continuity.
\end{proof}

Let $F$ have at least one edge, and let $\ell$ be a leaf of $F$ with neighbor $k$. As $\{k,\ell\}$ is the only edge of $F$ at $\ell$, all other edge differences of $F$ are determined by the directions of the users other than $\ell$.

For (a), we proceed by induction on $|E(F)|$, the case without edges being trivial. By the induction hypothesis, the other edge differences are linearly independent a.s. They are fewer than $d$, so some nonzero $\bu\in U$ is orthogonal to all of them. If $\bee_{k,\ell}$ were in their span, then $\langle\bee_{k,\ell},\bu\rangle=0$, i.e., $\langle\bp_\ell,\bu\rangle=\langle\bp_k,\bu\rangle$, which has probability zero by Lemma~\ref{lemma:circle}. Hence all columns of $\bE_F$ are linearly independent a.s.

For (b), fix $\bq\in(\bbZ+\tfrac12)^{d+1}$; as there are countably many such $\bq$, it suffices to show that $\bE_F^\top\bt=\bq$ has no solution a.s. By (a), the other $d$ edge differences are linearly independent a.s., hence form a basis of $U$. Their $d$ equations thus determine a unique $\bt\in U$, which is nonzero because the prescribed values are. The remaining equation prescribes $\langle\bee_{k,\ell},\bt\rangle$, hence $\langle\bp_\ell,\bt\rangle$, which has probability zero by Lemma~\ref{lemma:circle}.

\section{}
\label{Vitoria}

Fix a generator $\bt$ for $G$, and label vertex $k$ with
$\langle\bp_k,\bt\rangle\pmod 1$. Two vertices are adjacent if and
only if their labels differ by $\frac{1}{2}\pmod 1$.
In a nontrivial connected component, choose a vertex with label $\alpha$.
Following the paths from that vertex shows that every label is either
$\alpha$ or $\alpha+\frac{1}{2}\pmod 1$.
Vertices with the same label are not adjacent, while every pair with
different labels is adjacent. Thus, the two label classes form a
complete bipartite graph.

\section{}
\label{app:cover_family}

($\Rightarrow$) Suppose $G\in\cG$ is generated by $\bt\in U$.
Lemma~\ref{lemma:no_odd_cycle} gives~(i). Every edge of a spanning
forest $F$ of $G$ satisfies \eqref{general_position} for this $\bt$,
so Def.~\ref{def:general_position} gives $|E(F)|\le d$, which is~(ii).

($\Leftarrow$) Suppose $G$ satisfies~(i) and~(ii), and let $F$ be a
spanning forest, so that $|E(F)|\le d$.
Assign labels $0$ and $1$ to the two parts of each nontrivial
component, and write $\lambda(k)$ for the label of vertex $k$.
Isolated vertices may be assigned label $0$.
By Lemma~\ref{lemma:forest_independence}, the system
\begin{align}
    \langle\bee_{k,\ell},\bt\rangle
    =\frac{\lambda(\ell)-\lambda(k)}2
    \qquad \{k,\ell\}\in E(F),\quad k<\ell,
    \label{binary_forest_system}
\end{align}
has an affine solution set $A_F\subset U$ of dimension $d-|E(F)|$.
For vertices $k$ and $\ell$ in the same component, summing along
their unique path in $F$ gives
\begin{align}
    \langle\bee_{k,\ell},\bt\rangle
    =\frac{\lambda(\ell)-\lambda(k)}2
    \qquad \bt\in A_F.
\end{align}
This difference is $\pm1/2$ for vertices in opposite parts and zero
for vertices in the same part. Thus, within each component, every
$\bt\in A_F$ generates exactly the edges of $G$.

It remains to avoid edges between different components, including
isolated vertices. For such a pair $\{k,\ell\}$,
$F\cup\{\{k,\ell\}\}$ is a forest, and $\{k,\ell\}$ becomes an edge
exactly for  $\bt\in A_F$ satisfying \eqref{general_position} on
this forest. If $|E(F)|=d$, there is no such $\bt$ by general position.
If $|E(F)|<d$, Lemma~\ref{lemma:forest_independence} applied to this
forest makes $\bt\mapsto\langle\bee_{k,\ell},\bt\rangle$ nonconstant on
$A_F$, so such $\bt$ lie on countably many parallel hyperplanes of
$A_F$, a set of measure zero in $A_F$. In either case, some
$\bt\in A_F$ avoids all of the finitely many such pairs and thus
generates exactly $G$, proving $G\in\cG$.

\section{}
\label{app:covering_number}

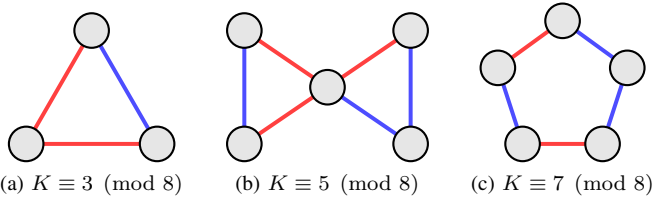
\begin{figure}[t]
    \centering

    \subfloat[$K\equiv 3 \pmod 8$\label{fig:residual-k3}]{
    \begin{tikzpicture}[
        vertex/.style={circle, draw=black, thick, fill=black!10, minimum size=13pt, inner sep=0pt},
        blockA/.style={line width=1.5pt, red!75},
        blockB/.style={line width=1.5pt, blue!70},
    ]
        \coordinate (a1) at (90:1cm);
        \coordinate (a2) at (210:1cm);
        \coordinate (a3) at (330:1cm);

        \draw[blockA] (a1) -- (a2);
        \draw[blockA] (a2) -- (a3);

        \draw[blockB] (a1) -- (a3);

        \foreach \i/\p in {1/a1,2/a2,3/a3}
            \node[vertex] at (\p) {};
    \end{tikzpicture}
    }
    \hspace{3mm}
    \subfloat[$K\equiv 5 \pmod 8$\label{fig:residual-k5}]{
    \begin{tikzpicture}[
        vertex/.style={circle, draw=black, thick, fill=black!10, minimum size=13pt, inner sep=0pt},
        blockA/.style={line width=1.5pt, red!75},
        blockB/.style={line width=1.5pt, blue!70},
    ]
        \coordinate (b1) at (0,0);
        \coordinate (b2) at (-1.1,0.75);
        \coordinate (b3) at (-1.1,-0.75);
        \coordinate (b4) at (1.1,0.75);
        \coordinate (b5) at (1.1,-0.75);

        \draw[blockA] (b1) -- (b2);
        \draw[blockA] (b1) -- (b3);
        \draw[blockA] (b1) -- (b4);

        \draw[blockB] (b1) -- (b5);
        \draw[blockB] (b5) -- (b4);
        \draw[blockB] (b2) -- (b3);

        \foreach \i/\p in {1/b1,2/b2,3/b3,4/b4,5/b5}
            \node[vertex] at (\p) {};
    \end{tikzpicture}
    }
    \hspace{3mm}
    \subfloat[$K\equiv 7 \pmod 8$\label{fig:residual-k7}]{
    \begin{tikzpicture}[
        vertex/.style={circle, draw=black, thick, fill=black!10, minimum size=13pt, inner sep=0pt},
        blockA/.style={line width=1.5pt, red!75},
        blockB/.style={line width=1.5pt, blue!70},
    ]
        \foreach \i in {1,...,5} {
            \coordinate (c\i) at ({90+(\i-1)*72}:0.9cm);
        }

        \draw[blockA] (c1) -- (c2);
        \draw[blockA] (c3) -- (c4);

        \draw[blockB] (c2) -- (c3);
        \draw[blockB] (c4) -- (c5);
        \draw[blockB] (c5) -- (c1);

        \foreach \i in {1,...,5}
            \node[vertex] at (c\i) {};
    \end{tikzpicture}
    }

    \caption{Case-by-case coverings of the residual graphs arising in the proof of Thm.~\ref{thm:covering_number}.}
    \label{fig:residual_coverings}
\end{figure}

Lemma~\ref{lemma:packing_number_cycles} supplies the required covers. For $K$
even, the uncovered perfect matching can be partitioned into
$\lceil K/6\rceil$ matchings of at most three edges. For
$K\equiv1\pmod8$, no edges remain uncovered; for
$K\equiv3,5,7\pmod8$, the residual graphs can each be covered by
two forests (Fig.~\ref{fig:residual_coverings}).

It remains to prove optimality. The bound in \eqref{trivial_bound}
suffices for $K\equiv1,5,7\pmod8$, leaving even $K$ and
$K\equiv3\pmod8$.
First suppose that $K$ is even. Counting edges gives
\begin{align}
    3\,\text{(\# forests)}+4\,\text{(\# 4-cycles)}
    \ge \tbinom{K}{2}.
    \label{edge_counting}
\end{align}
Let $d_F(v)$ and $d_C(v)$ be the sums of degrees of vertex $v$
across the forests and the 4-cycles, respectively. Since every
edge is covered,
\begin{align}
    d_F(v)+d_C(v)\ge K-1.
    \label{degree_counting}
\end{align}
If $d_F(v)=0$, the evenness of $d_C(v)$ implies $d_C(v)\ge K$;
otherwise, adding $d_F(v)\ge1$ to \eqref{degree_counting} gives
\begin{align}
    2d_F(v)+d_C(v)\ge K
    \label{improved_bound}
\end{align}
in either case. Each forest contributes at most six to
$\sum_v d_F(v)$, and each 4-cycle contributes eight to
$\sum_v d_C(v)$. Summing \eqref{improved_bound} gives
\begin{align}
    12\,\text{(\# forests)}+8\,\text{(\# 4-cycles)}
    \ge K^2.
    \label{parity_counting}
\end{align}
Adding $1/6$th of \eqref{edge_counting} and $1/24$th
of \eqref{parity_counting}, we obtain
\begin{align}
    \text{covering number}
    &\ge \tfrac16\tbinom{K}{2}+\tfrac1{24}K^2 \nonumber\\
    &=\tfrac{K(K-2)}8+\tfrac K6.
\end{align}
Since the first term is an integer for even $K$, this in turn gives
\begin{align}
    \text{covering number}
    \ge \tfrac{K(K-2)}8+\big\lceil\tfrac K6\big\rceil
\end{align}
as claimed.

Now suppose that $K\equiv3\pmod8$. Since
$\binom{K}{2}\equiv3\pmod4$, the claimed value exceeds
\eqref{trivial_bound} by one, so it suffices to rule out a cover
with $\lceil\binom{K}{2}/4\rceil$ graphs. As each forest has at
most three edges, such a cover has at most
\begin{align}
    4\big\lceil\tbinom{K}{2}/4\big\rceil-\text{(\# forests)}
    =\tbinom{K}{2}+1-\text{(\# forests)}
\end{align}
edges counted with multiplicity, so it contains at most one forest.
If it contains none, exactly one edge is covered twice and every
other edge once, so $d_C(v)=K$ at either endpoint $v$ of the
repeated edge. If it contains one, the forest has three edges and
the cover is edge-disjoint, so $d_C(v)=K-2$ at any leaf $v$ of the
forest. Both contradict the evenness of $d_C(v)$, as $K$ is odd.

\section{}
\label{app:pairing}

Let $C_1,C_2$ be the two $4$-cycles in Remark~\ref{remark:rotatable}, and let $\bt_1\in\cT_{C_1}$ and $\bt_2\in\cT_{C_2}$ be the generators spanning $V$. A generator $\beta_1\bt_1+\beta_2\bt_2\in V$ covers two edges with differences $\bee$ and $\bee'$ if $\langle\bee,\beta_1\bt_1+\beta_2\bt_2\rangle=\langle\bee',\beta_1\bt_1+\beta_2\bt_2\rangle=\tfrac12$, or equivalently,
\begin{align}
    \begin{bmatrix}
        \langle\bee,\bt_1\rangle & \langle\bee,\bt_2\rangle\\
        \langle\bee',\bt_1\rangle & \langle\bee',\bt_2\rangle
    \end{bmatrix}
    \begin{bmatrix}
        \beta_1 \\ \beta_2
    \end{bmatrix}
    =\begin{bmatrix}
        \frac{1}{2} \\ \frac{1}{2}
    \end{bmatrix}.
\end{align}
This linear system in $\beta_1$ and $\beta_2$ has a solution if its determinant
\begin{align}
    &\langle\bee,\bt_1\rangle\langle\bee',\bt_2\rangle-\langle\bee',\bt_1\rangle\langle\bee,\bt_2\rangle \nonumber\\
    &\quad= \big\langle\langle\bee,\bt_1\rangle\bee'-\langle\bee',\bt_1\rangle\bee,\,\bt_2\big\rangle
    \label{pairing_determinant}
\end{align}
is nonzero. To make the pairing within $V$ possible, it thus suffices to choose $\bt_1$ and $\bt_2$ so that \eqref{pairing_determinant} is nonzero for every two distinct edges. This can be done one generator at a time, as \eqref{pairing_determinant} is linear in $\bt_2$ once $\bt_1$ is fixed.

As noted after \eqref{lattice}, $\cT_{C_1}$ and $\cT_{C_2}$ are full-rank affine lattices, which no finite union of planes contains. Choose $\bt_1\in\cT_{C_1}$ off the planes $\langle\bee_{k,\ell},\bt\rangle=0$. For every two distinct edges, the vector $\langle\bee,\bt_1\rangle\bee'-\langle\bee',\bt_1\rangle\bee$ is nonzero, as $\bee$ and $\bee'$ are linearly independent by Lemma~\ref{lemma:forest_independence} and both coefficients are nonzero. Then, pick $\bt_2\in\cT_{C_2}$ off the planes orthogonal to these vectors, which makes \eqref{pairing_determinant} nonzero for every two distinct edges.

\section{}
\label{ApH}
Label vertex $k$ with
$\langle\bp_k,\bt\rangle\pmod 1$. In each connected component, all
labels belong to $\{\alpha+\frac{r}{b}:r=0,\ldots,b-1\}\pmod 1$ for some
$\alpha$. Vertices are adjacent exactly when their labels differ,
so the nonempty label classes form the parts.

\section{}
\label{ApI}
The proof follows that of Thm.~\ref{thm:cover_family} in
App.~\ref{app:cover_family}, using Lemma~\ref{lemma:multipartite} and
assigning distinct labels from $\{0,\ldots,b-1\}$ to the parts of each
component. Replace the half-differences of the binary labels by
$\frac{\lambda(\ell)-\lambda(k)}{b}$. Def.~\ref{def:general_position_radix}
supplies the forest solutions and excludes additional edges in the
same way.

\section{}
\label{ApJ}
Noncollinearity renders $\bee_{1,2}$ and $\bee_{1,3}$ independent, so
choose $\bt\in U$ with
$\langle\bee_{1,2},\bt\rangle=\frac13$ and
$\langle\bee_{1,3},\bt\rangle=\frac23$.
Then $\langle\bee_{2,3},\bt\rangle=\frac13$, and all three pairs satisfy
\eqref{radix_null_condition} with $b=3$.

\section{}
\label{app:counting_covers}

The number of covers with size $M$ is
\begin{align}
    \bigg|\bigg\{\{G_1,\ldots,G_M\}\subset\cG: \bigcup_m E(G_m) = E(H) \bigg\}\bigg|,
\end{align}
and its computation can be simplified through some tricks.

\begin{proposition}
\label{prop:inclusion_exclusion}
Let $f$ and $g$ be functions defined over the power set of $E(H)$, given by
\begin{align}
    f(T) &= \bigg|\bigg\{\{G_1,\ldots,G_M\}\subset\cG: \bigcup_m E(G_m) = T \bigg\}\bigg|\\
    g(S) &= \bigg|\bigg\{\{G_1,\ldots,G_M\}\subset\cG: \bigcup_m E(G_m) \subset S \bigg\}\bigg|,
\end{align}
where $T,S \subset E(H)$.
The number of the covers of size $M$ is
\begin{align}
    f\big(E(H)\big) = \sum_{T\subset E(H)} (-1)^{|E(H)|-|T|} g(T). \label{number_of_covers}
\end{align}
\end{proposition}
\begin{proof}
It trivially holds that
\begin{align}
    g(S) = \sum_{T\subset S} f(T).
\end{align}
Applying \cite[Thm. 7.6]{bona2017walk}, we have that
\begin{align}
    f(S) = \sum_{T\subset S} (-1)^{|S\setminus T|} g(T).
\end{align}
Plugging in $S=E(H)$ concludes the proof.
\end{proof}


The utility of Prop. \ref{prop:inclusion_exclusion} is that $g(T)$ is easy to evaluate:
\begin{align}
    g(T) &= \big|\big\{\{G_1,\ldots,G_M\}\subset\cG: E(G_m) \subset T \text{ for all }m \big\}\big| \nonumber\\
    &= \binom{|\{G\in\cG: E(G) \subset T\}|}{M}.
\end{align}
A direct evaluation of \eqref{number_of_covers} nonetheless remains out of reach, as the summation ranges over all $2^{\binom{K}{2}}$ labeled graphs on $K$ vertices---roughly $3.5\times 10^{13}$ for $K=10$. Instead, observe that $g(T)$ depends on $T$ only through its shape; that is, relabeling the vertices leaves $g(T)$ unchanged. It thus suffices to evaluate $g$ once for each distinct unlabeled graph on $K$ vertices and multiply by the number of vertex labelings that produce it. The number of unlabeled graphs on $K$ vertices (OEIS~A000088 \cite{oeis_a000088}) is only about $1.2\times 10^{7}$ for $K=10$; enumerating them with \texttt{nauty} \cite{mckay2014practical} renders the computation tractable, yielding
\begin{align*}
    6 &\text{ for }K=4\\
    95 &\text{ for }K=5\\
    105 &\text{ for }K=6\\
    1\,626\,660 &\text{ for }K=7\\
    9\,865\,742\,670 &\text{ for }K=8\\
    1\,643\,040 &\text{ for }K=9\\
    43\,957\,304\,088\,075 &\text{ for }K=10.
\end{align*}
For $K=9$, all graphs in the cover should be 4-cycles, resulting in a sharp drop in the number of covers; the complete graph on 9 vertices has 36 edges and its covering number is $9=\frac{36}{4}$ (recall Lemma \ref{lemma:packing_number_cycles} and Table \ref{table:covering_number}).

\section{}
\label{app:minimum_separation}

For Fig.~\ref{fig:min_separation}, the mean width is minimized over the minimum covers $\{G_1,\ldots,G_M\}$ and the positive odd integers $\{\kappa_m\}$, subject to the minimum separation constraint. Let $W$ be the mean width of an array satisfying the constraint, which exists by Sec.~\ref{sec:minsep}. The optimum satisfies $\frac12\sum_{m=1}^M\kappa_m\|\bt_{G_m}\|\le W$, which leaves finitely many $\{\kappa_m\}$ for each minimum cover. The constraint is checked for each.

\bibliographystyle{IEEEtran}
\bibliography{ref}

\end{document}